\documentclass[preprint,showpacs,preprintnumbers,amsmath,amssymb,prb]{revtex4-1}

\usepackage[dvipdfmx]{graphicx}
\usepackage{dcolumn}
\usepackage{bm}
\usepackage{mhchem} 
\usepackage{ascmac}
\usepackage{color} 
\usepackage[normalem]{ulem}
\begin{document}
%
\title{Photoluminescence decay of mobile carriers influenced by imperfect quenching at particle surfaces with subdiffusive spread}
\author{Ryuzi Katoh}
\affiliation{
College of Engineering, Nihon University, Koriyama, Fukushima 963-8642, Japan
}
\author{
Kazuhiko Seki
}
\email{k-seki@aist.go.jp}
\affiliation{GZR, National Institute of Advanced Industrial Science and Technology (AIST), Onogawa 16-1 AIST West, Ibaraki 305-8569, Japan
}
\date{\today}
\begin{abstract}
We recently presented a quantitative model to explain the particle-size dependence of photoluminescence (PL) quantum yields and revealed that exciton quenching is not diffusion controlled, but limited by surface reactions.
However, the exciton decay kinetics has not yet been analyzed using our theoretical model. 
Here, we study kinetic aspects of the model and show that it should be extended to take into account subdiffusion rather than normal diffusion to maintain consistency with the observed complex decay kinetics; we also show that the PL decay kinetics is nonexponential even when the PL quenching is limited by surface reactions under subdiffusion.
Our theoretical analysis of the PL quantum yield and the PL decay kinetics provides a comprehensive picture of mobile charge carriers, immobile polarons, and self-trapped excitons. 
\end{abstract}

\maketitle
\section{Introduction}
TiO$_2$ crystals have attracted attention as photocatalytic materials over the past several decades following the discovery of photoinduced splitting of water into hydrogen and oxygen, which was first demonstrated using TiO$_2$ electrodes.\cite{FUJISHIMA_72}
Research on photoinduced water splitting by semiconductors has evolved to include increasing the energy conversion efficiency by doping TiO$_2$\cite{KUMARAVEL_19} or using semiconductors with a narrower bandgap\cite{Pihosh_23}; it has also expanded to include scaling up by using particulate photocatalysts rather than electrodes.\cite{Takata_20,Nishiyama_21}
The concept of photoenergy conversion using TiO$_2$ crystals has been expanded to include the photocatalytic decomposition of pollutants, wastes, and bacteria, as well as applications in dye-sensitized solar cells.\cite{Wang_97,FUJISHIMA_00,FUJISHIMA_08,ORegan_91,PELAEZ_12}

The principles and mechanisms of TiO$_2$-based photocatalysis have been also explored for decades.\cite{Linsebigler_95,Schneider_14,WEN_15}
Transient absorption spectroscopy (TAS) measurements carried out using pulsed photoexcitation are a powerful method to characterize the decay of photoexcited charge carriers via their recombination and their transfer into and out of immobile states.\cite{Yoshihara_04,Tamaki_06,YAMAKATA_07}
The time-resolved microwave conductivity (TRMC) technique has also been used to measure the conductivity of charge carriers in TiO$_2$.\cite{Warman_84,Colbeau-Justin_03,Kroeze_04,Fravventura_13,Saeki_14,Nakajima_15}
The TAS and TRMC measurements have both provided fundamental insights into carrier relaxation processes of excited electrons and holes in TiO$_2$. 

In addition to the TAS methods, photoluminescence (PL) provides alternative and additional information about the nature and kinetics of photoexcited charge carriers and excitons in TiO$_2$.\cite{FUJIHARA_00,Harada_07,Cavigli_09,Wang_10,Yamada_12,Pallotti_17,Gallart_18,Vequizo_18,Bruninghoff_19,Krivobok_20,Katoh_22,Katoh_24}
PL spectra of TiO$_2$ photocatalysts have been acquired using both continuous photoexcitation and pulsed photoexcitation.\cite{FUJIHARA_00,Harada_07,Cavigli_09,Wang_10,Yamada_12,Pallotti_17,Gallart_18,Vequizo_18,Bruninghoff_19,Krivobok_20,Katoh_22,Katoh_24}
Broad emission spectra have been observed for various TiO$_2$ particles; the spectral shape depends on the form of TiO$_2$ (i.e., anatase, rutile, or brookite) and is time-dependent.\cite{Vequizo_18,Bruninghoff_19,Katoh_22,Katoh_24}
The PL intensity of TiO$_2$ increases at low temperatures.\cite{Harada_07,Cavigli_09,Wang_10,Krivobok_20} 
However, the origins of the differences in the PL spectra and their time-dependence are still not fully understood. 

The PL quantum yields of anatase and rutile TiO$_2$ particles have recently been measured.\cite{Katoh_22,Katoh_24}
The PL quantum yields of nanosized particles were found to be low ($<10^{-4}$) and dependent on the particle size. 
For anatase TiO$_2$, the quantum yields are proportional to the particle diameter.\cite{Katoh_22,Katoh_24}
The observed increase of the PL quantum yield with increasing particle size indicates that PL is emitted mainly from the bulk and that the quenching centers are located on the surfaces of TiO$_2$ particles.\cite{Katoh_22,Katoh_24} 
To gain a comprehensive understanding of the origin of the PL emission of TiO$_2$, we investigated the PL decay kinetics for anatase and rutile TiO$_2$ particles whose quantum yields were known.\cite{Katoh_24}
The measured PL decay of anatase TiO$_2$ was nonexponential and complex, and the results were difficult to rationalize.  
At the nanosecond timescale, the decay kinetics depends on the particle size; however, at the picosecond timescale, the decay kinetics is independent of particle size. 
We have also proposed a quantitative model to understand the particle-size dependence of the PL quantum yields and have found that exciton quenching is not diffusion controlled but limited by surface reactions; the partial surface quenching of excitons is consistent with the observation that a long-lived PL component remained over a microsecond timescale. \cite{Katoh_24} 
However, the exciton decay kinetics has not been analyzed using this theoretical model. 

In the present study, we investigate kinetic aspects of our previously proposed model and show that, to maintain consistency with the observed complex decay kinetics, the model should be extended to take into account subdiffusion rather than normal diffusion. 
Subdiffusive carrier transport as a result of the decrease in diffusivity over time has been reported for quantum-dot solids and semiconductors.\cite{Akselrod_14,Seitz_20,Ginsberg_20}
The mean square displacement of carriers obeys $\langle r^2(t) \rangle =2 d D_\alpha t^\alpha$, where $\alpha<1$, $\langle \cdots \rangle$ indicates the ensemble average and $d$ is the dimensionality for the region of diffusion. 
Parameter $D_\alpha$ is the diffusion coefficient of subdiffusion, whose dimensionality ([Length]$^2$/[Time]$^\alpha$) differs from that of the diffusion constant of normal diffusion  for $\alpha<1$.
We confirm that the PL quantum yield is proportional to the particle size even under subdiffusion when excitons are partially quenched at the particle surfaces. 
The PL decay can be approximated using the one--parameter Mittag--Leffler function, which can be further simplified using the stretched exponential decay function and the power-law decay function. 
The characteristic time constant which divides the time regime of stretched exponential decay by that of the power-law decay is shown to depend on the particle size.
These results provide a theoretical framework for analyzing complex PL decay kinetics, which can be approximated by either the stretched exponential decay function or the power-law decay function.\cite{FUJIHARA_00,Katoh_24}
The theoretical results are used to analyze the PL decay kinetics at the nanosecond timescale, where the decay is particle-size dependent.

The size dependence of PL decay inevitably requires a through study of diffusive migration of excitons. 
So far, a limited number of models are available for studying the relation between PL kinetics and diffusive properties of excitons in confined inorganic materials. 
\cite{TANG_93,Tang_95,Gallart_18,Katoh_22,Katoh_24,Bruninghoff_19,Kurilovich_20,Kurilovich_22,Kurilovich_24,Wietek_24}
Here, we introduce carriers that diffuse inside the particle, emit PL (or form excitons to emit PL) in the bulk, and are quenched at the particle surface. 
In the simplest model, the carriers are mobile excitons. 
When the formation of self-trapped excitons with a large Stokes shift is considered, self-trapped excitons might become mobile by dissociating into a mobile charge carrier and an immobile polaron.\cite{TANG_93,Tang_95,Gallart_18,Bruninghoff_19}
Broad emission bands suggest that self-trapped excitons might be composed of a strongly bound polaron and a weakly bound charge carrier with various configurations.\cite{TANG_93,Tang_95,Gallart_18}
When a self-trapped exciton dissociates into a mobile charge carrier and an immobile polaron, the carriers in our model are mobile charge carriers. 

Carriers with a sufficiently long bulk lifetime can be spread by diffusion and occasionally quenched at the particle surface; the carriers can escape from quenching at the particle surface and diffuse again inside the particle. 
Accordingly, PL decays slowly as a result of carriers occasionally visiting the particle surface, including repeated visits.\cite{Katoh_24}
When the diffusion length exceeds the particle size and carriers are perfectly quenched at the particle surface, the quantum yield is proportional to the square of the particle size.\cite{Katoh_24}
The measured quantum yield is proportional to the particle size, supporting our model of surface partial quenching of carriers.\cite{Katoh_24}

PL decay occurs over a wide range of timescales.\cite{Vequizo_18,Bruninghoff_19,Krivobok_20,Katoh_22,Katoh_24}
In previous studies, PL with a lifetime of hundreds of nanoseconds was observed, and the spectra were observed to slowly shift to longer wavelengths over time.\cite{Vequizo_18,Bruninghoff_19,Katoh_24}
The presence of long-lived carriers inside the particle further supports our model of partial quenching at the particle surface.\cite{Katoh_22,Katoh_24}

In Sec. \ref{sec:model}, we describe the model. 
In Sec. \ref{sec:Analysis}, 
we present the analysis of PL kinetics.
Theoretical results are presented in Secs. \ref{sec:Normaldiff}-\ref{sec:QY}.
The fitting function in Sec. \ref{sec:Analysis} is justified by comparison with the numerical exact solution in Sec. \ref{sec:Discussion}.
The conclusion is given in Sec. \ref{sec:Conclusion}.

\section{Model}
\label{sec:model}

\begin{figure}[h]
\begin{center}
\includegraphics[width=0.8\textwidth]{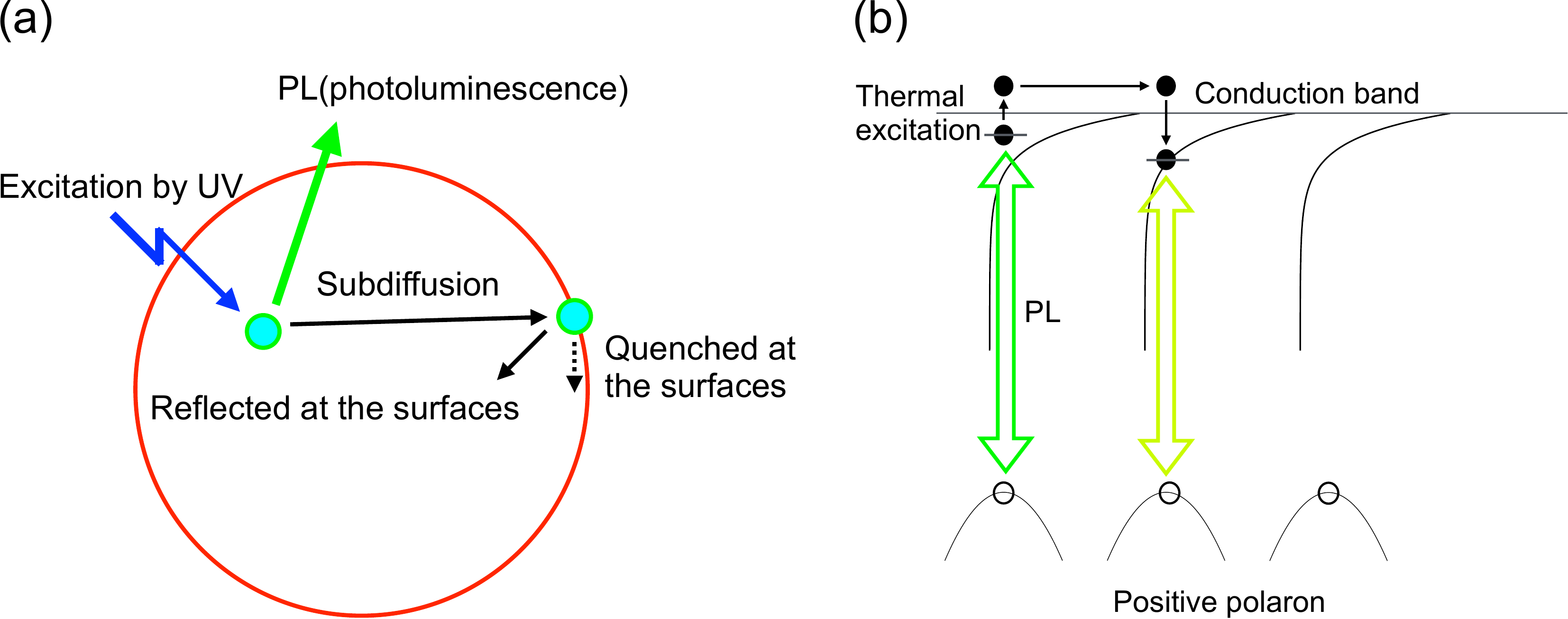}
\end{center}
\caption{(Color online) (a) Schematic of the model. (b) Schematic of subdiffusion process. 
A self-trapped exciton executes the following processes: dissociation of a self-trapped exciton to a mobile charge carrier, leaving behind an immobile polaron; migration of a mobile charge carrier; and association of the mobile charge carrier with one of the immobile polarons to form a self-trapped exciton again.
The PL process competes with the thermal excitation process. 
A self-trapped exciton is composed of a mobile charge carrier Coulombically bound to an immobile polaron with various configurations that yield an exponentially distributed binding energy. 
As time proceeds, the distribution of self--trapped excitons shifts towards the stronger binding energy. 
This model is consistent with the observed red-shift of the PL emission wavelength over time after pulsed excitation. 
}
\label{Fig:model}
\end{figure}

We previously presented a theoretical model to explain the size dependence of the PL quantum yield, where the PL arises from excitons in the bulk of the particle and the quenching centers of PL are located on the particle surfaces.\cite{Katoh_24}
The rate constant for quenching on the surface of the particle and the diffusion length of excitons were introduced.\cite{Katoh_24}
The diffusion length is defined as the square root of the diffusion constant multiplied by the natural lifetime (bulk lifetime) of excitons in sufficiently large samples where the influence of boundaries can be ignored.
We have shown that the PL quantum yield is proportional to the particle size when the diffusion length exceeds the particle size and that the PL decays by being partially quenched at the particle surface.\cite{Katoh_24}

In the present study, we investigate the PL kinetics using the same theoretical model developed to study the particle-size dependence of the PL quantum yield. 
The model is schematically shown in Fig. \ref{Fig:model} (a).
As we will show later, highly nonexponential decay of PL cannot be modeled if we assume normal diffusion in the bulk.
For consistency with the observed complex decay kinetics, the model is extended to consider subdiffusion. [Fig. \ref{Fig:model} (b)] 
We will show that the quantum yield can still be proportional to the particle size if the diffusion length of carriers, which can be properly defined for the case of subdiffusion, exceeds the particle size and carriers are partially quenched at the particle surface. 
We will also show that the PL kinetics can be highly nonexponential because of subdiffusion. 

In this model, we consider a spherical particle with radius $R$. 
We consider the case where radius $R$ is smaller than the absorption depth of incident light and assume a uniform initial distribution of carriers within the spherical particle; 
the carriers are generated uniformly inside the spherical particle by a pulsed light excitation. 
The initial density of the carriers is denoted by $p_0$, which might be proportional to the excitation light-intensity; 
the total number of photo-generated carriers inside the spherical particle is 
$4 \pi R^3 p_0/3$.
We denote the distribution of carriers at time $t$ by $p(r,t)$, where $r$ is the distance from the center of the particle. 
We have $p(r,0)=p_0$ by assuming the uniform initial distribution. 
The PL rate (PL intensity) can be obtained from 
$I_{\rm PL}(t)=4\pi \int_0^R dr r^2 k_{\rm r} p (r,t)$, where $ k_{\rm r}$ is the radiative-decay rate constant. 
The normalized PL intensity can be expressed as 
$I_{\rm PL}^{\rm (n)}(t)=I_{\rm PL}(t)/I_{\rm PL}(0)$, where we have $I_{\rm PL}(0)=4 \pi R^3 k_{\rm r} p_0/3$. 
We introduce the carrier-deactivation rate constant ($k_{\rm t}$), where the total deactivation rate is given by the sum of the radiative-decay rate constant ($k_{\rm r}$) and the nonradiative-decay rate constant ($k_{\rm nr}$), {\it i.e.}, $k_{\rm t}=k_{\rm r}+k_{\rm nr}$. 
Parameter $k_{\rm t}$ can be regarded as the natural decay rate constant unaffected by the particle boundary conditions.
We consider the PL decay induced by quenching at the particle surface. 

In the simplest model, carriers are excitons, as already stated. 
When an exciton dissociates into a mobile charge carrier, leaving behind an immobile polaron,\cite{TANG_93,Gallart_18,Bruninghoff_19} 
the mobile charge carrier can associate with any of the immobile polarons to form an emissive exciton again. 
The mobile charge carrier can be either an electron or a hole. 
For brevity, we assume below that electrons are mobile. 
The self-trapped exciton can be regarded as a random walker stepping on the locations of immobile polarons, thereby repeatedly executing the following processes: dissociation of a self-trapped exciton to a mobile charge carrier, leaving behind an immobile polaron; migration of a mobile charge carrier; and association of the mobile charge carrier with one of the immobile polarons to form a self-trapped exciton again. [Fig. \ref{Fig:model} (b)] 
(Note that strongly bound positive polarons located inside the particle might remain inside the particle even when the particle surface is exposed to hole scavengers. 
In this case, only the surface quenching sites are scavenged.) 
Given that TiO$_2$ crystals are intrinsically n-type semiconductors, positive polarons can be generated by photoexcitation alone.\cite{Carey_21}
If a sufficient density of immobile polarons is generated by photoexcitation, then $k_{\rm t}$ is related to the natural decay rate constant of self-trapped excitons rather than to the natural decay rate constant of mobile charge carriers. 
If the occasional dissociation of excitons is considered, then $k_{\rm t}$ is smaller than the natural decay rate constant of self-trapped excitons. 

Thus far, the model has been described in terms of excitons because the model will be used to analyze the PL from self-trapped excitons in anatase TiO$_2$.\cite{TANG_93,Tang_95,Gallart_18,Katoh_22,Katoh_24} 
The theoretical model itself, however, can also be used to study the decay of minority charge carriers in semiconductors through recombination with majority charge carriers and quenching at the particle surfaces, where $k_{\rm t}$ is the inverse of the natural decay rate constant of minority charge carriers. 
PL can be emitted when a minority charge carrier recombines with a majority charge carrier. 
In this case, the carriers are minority charge carriers. 
The majority charge carriers and minority charge carriers for n-type semiconductors are electrons and holes, respectively. 
Parameter $k_{\rm t}$ can be given by the rate constant of the recombination of the minority charge carriers and the majority charge carriers multiplied by the dark (equilibrium) density of the majority charge carriers. 
The minority charge carriers can be either quenched or reflected at the particle surfaces. 
In the latter case, the minority charge carriers migrate again inside the particle.  
The internal quantum yield of photocatalysts has been investigated using this theoretical model.\cite{Nandal_21}

\section{Analysis of PL kinetics}
\label{sec:Analysis}

Three anatase TiO$_2$ powders with particle diameters of 7, 25, and 200 nm were purchased from commercial sources; these powders are denoted as A-7 [Ishihara  (ST01)], A-25 [Sigma-Aldrich (637254)], and A-200 [Ishihara (ST41)], respectively. 
Measurements were performed at $295$ K under ambient air conditions immediately after the materials were retrieved from their containers. 
The experimental PL measurement methods are described elsewhere.\cite{Katoh_24}
The PL quantum yield data as a function of the particle diameter were previously analyzed theoretically by assuming normal diffusion.\cite{Katoh_24} 
For the purpose of self-consistency, the measured PL quantum yield is shown as a function of particle diameter in Appendix A (Fig. \ref{Fig:QE}); the PL quantum yield is proportional to $R$ (partial quenching case) rather than $R^2$ (perfect quenching case), as reported previously.\cite{Katoh_24} 
However, according to our theoretical model, the highly nonexponential PL decay observed in Ref. \onlinecite{Katoh_24} indicates subdiffusion rather than normal diffusion  
as shown later in details.
 We will also show that the PL quantum yield can be proportional to the particle size even under subdiffusion; the interpretation of partial quenching at the particle surfaces remains unchanged under subdiffusion. 
(See Table \ref{table:2}.)
Below, to analyze PL decay kinetics under subdiffusion, we assume partial quenching at the particle surface. 

The partial quenching at the particle surface is characterized by the rate coefficient denoted by $k_{{\rm q}\alpha}$. 
The quenching of carriers at the particle surface occurs in competition to subdiffusive escape from the surface to the bulk (inside the particle).
As a result,  the dimensionality of $k_{{\rm q}\alpha}$ is [length]/[time]$^\alpha$, where $\alpha$ indicates the parameter of sub-diffusion expressed by 
$\langle r^2(t) \rangle =2 d D_\alpha t^\alpha$. [See explanation below Eq. (\ref{eq:BCds}).] 
We consider the case that carriers are generated uniformly inside the particle by pulsed excitation. 
The carriers execute subdiffusion inside the particle and are partially quenched at the particle surface. 
We will show that 
the PL decay can be expressed by the one--parameter Mittag--Leffler function [Eq. (\ref{eq:PLsappr1ds})] 
when the particle size is smaller than the diffusion length. 
Here, we present the results of analyzing the PL decay using Eq. (\ref{eq:PLsappr1ds}), before describing the derivation of this equation.  

\begin{figure}[h]
\begin{center}
\includegraphics[width=0.5\textwidth]{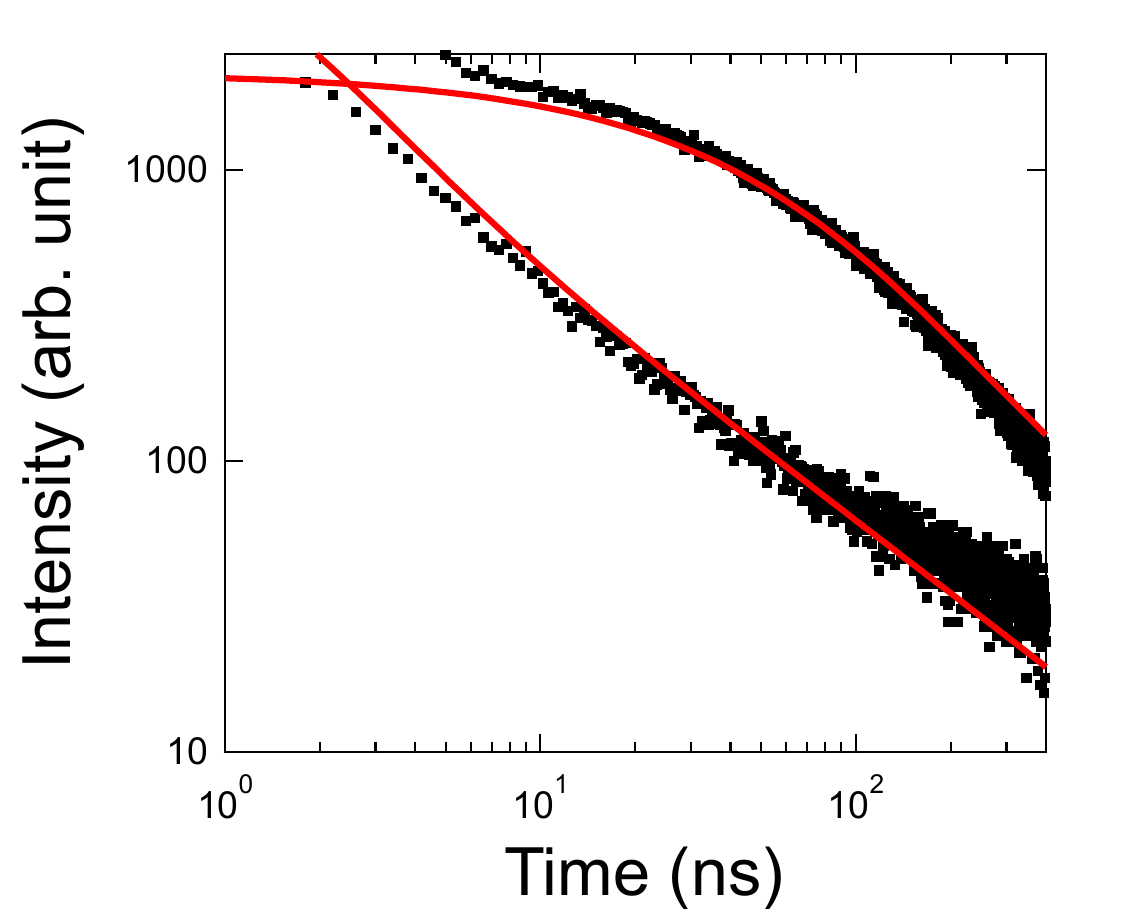}
\end{center}
\caption{(Color online) 
The upper dots and the lower dots indicate the PL decay profile of A-200 ($R=100$ nm) and that of A--7 ($R=3.5$ nm), respectively, excited at $355$ nm and observed at $600$ nm.\cite{Katoh_24}
The red lines are curves fitted using the one--parameter Mittag--Leffler function [Eq. (\ref{eq:PLsappr1ds})] with $k_{\rm t}=10^5$ s$^{-1}$, $\alpha=0.8$, $k_{{\rm q}\alpha}=0.0214$ ms$^{-\alpha}$, and the value of $D_\alpha$ satisfying $R\sqrt{k_{\rm t}^\alpha/D_\alpha} =0.01$ with $R=100$ nm. 
 }
\label{Fig:A200}
\end{figure}
In Fig. \ref{Fig:A200}, we applied the one--parameter Mittag--Leffler function [Eq. (\ref{eq:PLsappr1ds})] to fit the A-200 ($R=100$nm) data to the model and found $k_{\rm t}=10^5$ s$^{-1}$, $\alpha=0.8$, and $k_{{\rm q}\alpha}=0.0214$ m s$^{-\alpha}$. 
In Fig. \ref{Fig:A200}, we also show that the data for A-7 ($R=3.5$ nm) are correlated with the theoretical line obtained from Eq. (\ref{eq:PLsappr1ds}) when the size is reduced from $R=100$ nm to $R=3.5$ nm while the other parameter values are maintained ($k_{\rm t}=10^5$ s$^{-1}$, $\alpha=0.8$, and $k_{{\rm q}\alpha}=0.0214$ m s$^{-\alpha}$).
By retaining the exponent of subdiffusion ($\alpha=0.8$) and the surface extraction rate coefficient ($k_{{\rm q}\alpha}=0.0214$ ms$^{-\alpha}$), we can interpret the results as the particle size being reduced from $R=100$ nm to $R=3.5$ nm; the bulk and surface properties of A-200 and A-7 are the same. 
We can define the decay time constant $\tau$ using $E_{\alpha} \left[-(t/\tau)^\alpha \right]$ in Eq. (\ref{eq:PLsappr1ds}), {\it i.e.}, $\tau=\left[ R/(3k_{{\rm q}\alpha}) \right]^{1/\alpha}$. 
The time scale of A-200 ($R=100$ nm) decay can be estimated as $\tau=55$ ns using $k_{{\rm q}\alpha}=0.0214$ m s$^{-\alpha}$ and $\alpha=0.8$. 

\begin{figure}[h]
\begin{center}
\includegraphics[width=0.5\textwidth]{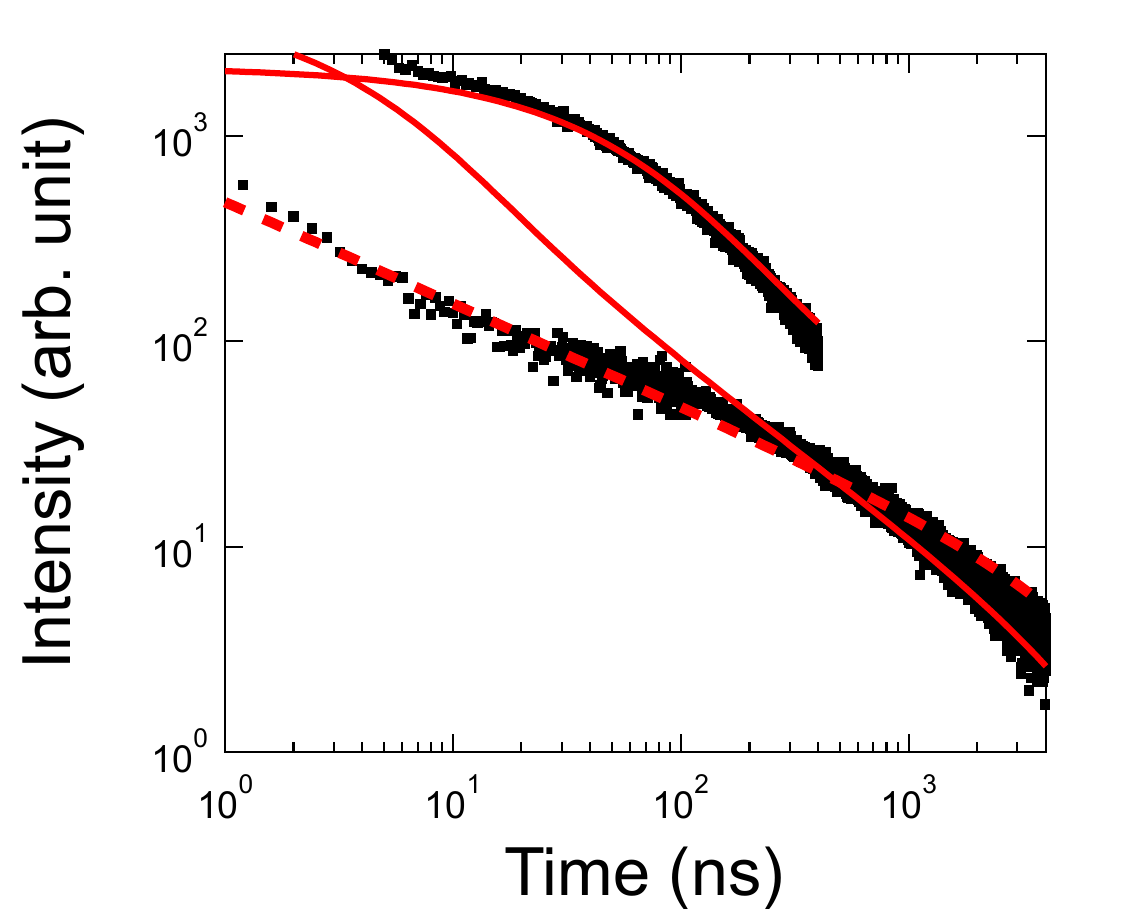}
\end{center}
\caption{(Color online) 
The upper dots and the lower dots indicate the PL decay profile of A-200 ($R=100$ nm) and that of A-25 ($R=12.5$ nm), respectively, excited at $355$ nm and observed at $600$ nm.\cite{Katoh_24}
The red solid lines are fitted curves for the one--parameter Mittag--Leffler function [Eq. (\ref{eq:PLsappr1ds})] using the same parameters as those corresponding to Fig. \ref{Fig:A200} ($k_{\rm t}=10^5$ s$^{-1}$, $\alpha=0.8$, and $k_{{\rm q}\alpha}=0.0214$ m s$^{-\alpha}$).  
The dashed line is obtained using $k_{\rm t}=10^5$ s$^{-1}$, $\alpha=0.5$, and $k_{{\rm q}\alpha}=6.77\times 10^{-4}$ m s$^{-\alpha}$, which is determined to ensure the proportional relation between the PL quantum yield and the particle size ($k_{\rm t}^{0.8}/0.0214=k_{\rm t}^{\alpha}/k_{{\rm q}\alpha}$), where the value of $\alpha$ is changed. 
All theoretical lines are obtained by assuming the value of $D_\alpha$ satisfying $R\sqrt{k_{\rm t}^\alpha/D_\alpha} =0.01$ with $R=100$ nm, as in Fig. \ref{Fig:A200}. 
}
\label{Fig:A25}
\end{figure}

In Fig. \ref{Fig:A25}, we show the theoretical line (the lower solid (red) line) obtained from Eq. (\ref{eq:PLsappr1ds}) by reducing the size from $R=100$ nm to $R=12.5$ nm while keeping the same values used to draw Fig. \ref{Fig:A200} for the other parameters ($k_{\rm t}=10^5$ s$^{-1}$, $\alpha=0.8$, and $k_{{\rm q}\alpha}=0.0214$ ms$^{-\alpha}$). 
The data for A-25 ($R=12.5$ nm) deviate from the theoretical line (the lower solid (red) line) in the time regime below $100$ ns, unlike the cases of A-7 ($R=3.5$ nm) and A-200 ($R=100$ nm). 
The better agreement in the time regime below $100$ ns can be attained by changing $k_{{\rm q}\alpha}$. 
However, if $k_{{\rm q}\alpha}$ alone is changed while keeping $\alpha$ unaltered, the PL quantum yield is theoretically no longer proportional to $R$ according to Eq. (\ref{eq:PhiPLsds_1}); $\alpha$ is the exponent of subdiffusion inside the particle. 
If we assume that $\alpha$ can be changed when the particle size is changed, then the proportional relation between the PL quantum yield and the size still holds if $k_{\rm t}^{\alpha}/k_{{\rm q}\alpha}$ in the expression of $\Phi_{\rm PL}\approx (k_{\rm r}/k_{\rm t}) k_{\rm t}^{\alpha}R /(3k_{{\rm q}\alpha})$ [Eq. (\ref{eq:PhiPLsds_1})] is retained. 

We change $\alpha$ and $k_{{\rm q}\alpha}$ to satisfy $k_{\rm t}^{0.8}/0.0214=k_{\rm t}^{\alpha}/k_{{\rm q}\alpha}$ to maintain the proportional relation between the PL quantum yield and the size, where $k_{\rm t}=10^5$ s$^{-1}$ is assumed using the data for PL decay over a period of 1 $\mu$s. 
With this approach, the dashed (red) line in Fig. \ref{Fig:A25} is obtained with $\alpha=0.5$, and $k_{{\rm q}\alpha}=6.77\times 10^{-4}$ m s$^{-\alpha}$; the data for A-25 ($R=12.5$ nm) are correlated with the dashed line. 
The $\alpha$ value of 0.5 is consistent with the value previously obtained by fitting to the power-law decay function\cite{Katoh_24} and indicates that the exponent of the bulk subdiffusion of A-25 is lower than the exponent ($\alpha=0.8$) of A-200 and A-7. 
The difference in bulk properties is attributable to the A-25 being purchased from a different company than the A-200 and A-7.
We selected anatase TiO$_2$ that is free from emissions possibly attributable to rutile TiO$_2$. 
A subtle difference in the electron trap distribution for commercially available TiO$_2$ has also been reported.\cite{Nitta_16}
Additional careful experimental studies on the size dependence of PL kinetics are needed; different values of $\alpha$ should not be obtained by changing the particle size, irrespective of whether the stretched exponential function or the power-law decay function is used.
Notably, the one--parameter Mittag--Leffler function can be approximated by either the stretched exponential function or the power-law decay function with the same exponent $\alpha$ [Eqs. (\ref{eq:MLexpand})--(\ref{eq:MLasymptotic})].
In the subsequent sections, we give detailed theoretical derivations. 

\section{Theory: Normal diffusion}
\label{sec:Normaldiff}

Denoting the diffusion constant as $D$, we solve the diffusion equation under the influence of the deactivation rate constant denoted by $k_{\rm t}$:
\begin{align}
\frac{\partial}{\partial t}  p (r,t)=D \nabla^2 p (r,t)-k_{\rm t} p (r,t),
\label{eq:diff}
\end{align}
where the Laplacian is expressed by assuming spherical symmetry, 
\begin{align}
\nabla^2= \frac{\partial^2}{\partial r^2} + \frac{2}{r} \frac{\partial}{\partial r}=\frac{1}{r^2} \frac{\partial}{\partial r} r^2 \frac{\partial}{\partial r}. 
\label{eq:laplace3d}
\end{align}
Equation (\ref{eq:diff}) represents the diffusion of carriers, where the natural decay rate constant is given by $k_{\rm t}$ inside the particle. 

\subsection{Partial quenching at the particle surface}
\label{sec:partialq}
Here, we assume that carriers are partially quenched at the particle surface,
\begin{align}
\left. - D \frac{\partial}{\partial r} p (r,t)\right|_{r=R}= k_{\rm q} p (R,t).
\label{eq:BC}
\end{align}
Note that $k_{\rm q}$ is independent of $R$ from the following consideration. 
Equation (\ref{eq:BC}) can be obtained by considering that the total probability current of carriers flowing to the surface of radius $R$ is obtained as 
$j=- 4 \pi R^2 D(\partial p (r,t)/\partial r) |_{r=R}$ 
and that the total current is equal to the deactivation rate at the total surface given by $4 \pi R^2 k_{\rm q} p (R,t)$, where $k_{\rm q}$ indicates the rate constant of surface reaction (at the unit surface area) between $R$ and $R+dr$ and the dimensionality of $k_{\rm q}$ is [length]/[time].
Therefore, $k_{\rm q}$ is the second-order rate constant. 
Parameter $k_{\rm q}$ is independent of $R$ because the influence of $R$ has already been taken into account by $4 \pi R^2$ in $4 \pi R^2 k_{\rm q} p (R,t)$. 
A similar consideration has been presented for lateral growth of a circular domain.\cite{SEKI_21,Seki_22}

As long as the current density of carriers at the surface is irreversibly quenched and the quenching rate is proportional to the carrier density at the surface, 
the boundary condition can be expressed by 
Eq. (\ref{eq:BC}) using the rate constant denoted by $k_{\rm q}$; 
the theoretical results follow from the boundary condition given by Eq. (\ref{eq:BC}) 
regardless of the origin of the surface quenching. 
Although various mechanism of PL quenching at the surface of anatase TiO$_2$ has been proposed, \cite{Vequizo_18,Bruninghoff_19,Krivobok_20,Katoh_22}
the boundary condition given by Eq. (\ref{eq:BC}) can be applicable as long as the surface quenching is irreversible. 
As an origin of PL quenching, 
the photoluminescence has been shown to be correlated with the exposure to 
alcohol; 
the photoluminescence is enhanced by alcohol exposure. \cite{Knorr_08,Vequizo_18}
By considering that alcohol reacts with holes on the particle surface, \cite{Tamaki_06}
the photoluminescence may be enhanced by reducing the surface trapped holes by alcohol exposure.   
\cite{Knorr_08,Vequizo_18}
Conversely, quenching of photoluminescence might be correlated with the concentration of surface trapped holes.  

We denote the Laplace transform of $f(r,t)$ by $\hat{f}(r,s)$. 
The inverse Laplace transformation is denoted by ${\cal L}^{-1} \left(\hat{f}(r,s)\right)$, which is equal to $f(r,t)$. 
We calculate the Laplace transform of the normalized PL intensity denoted by $\hat{I}_{\rm PL}^{\rm (n)} (s)$ as shown in Appendix B, 
where we have $I_{\rm PL}(0)=4\pi R^3 k_{\rm r} p_0/3$ by assuming uniform generation of excitons inside the spherical particle. 
We find for $I_{\rm PL}^{\rm (n)}(t)=I_{\rm PL}(t)/I_{\rm PL}(0)$, 

\begin{align}
\hat{I}_{\rm PL}^{\rm (n)}(s)
&= \frac{1}{s+k_{\rm t}}
\left[  1
 - 3 \frac{L_{\rm D}^2(s)}{R^2}
 \left(\frac{D}{k_{\rm q}R}+
  \left(\frac{R}{L_{\rm D}(s)}\coth\left( \frac{R }{L_{\rm D}(s)}\right)- 1 \right)^{-1}
 \right)^{-1}
\right] ,
\label{eq:PLs}
\end{align}
where $L_{\rm D}(s)$ is defined by  
\begin{align}
L_{\rm D}(s)=\sqrt{D/(s+k_{\rm t})}, 
\label{eq:LDs}
\end{align}
which reduces to the diffusion length in the limit of $s\rightarrow 0$. 

\subsubsection{Particle size smaller than the diffusion length}
\label{sec:sparticlesize}

First, we consider the case $R/L_{\rm D}(s=0)<1$, which can be expressed as $R\sqrt{k_{\rm t}/D} <1$; this condition applies when the particle size is sufficiently small and has been considered previously to study the size dependence of the PL quantum yield.\cite{Katoh_24}

The asymptotic kinetics can be obtained by assuming $R/L_{\rm D}(s)<1$. 
Using $x \coth(x)-1 \approx x^2/3-x^4/(45) +\cdots $ for $x<1$, we can approximate Eq. (\ref{eq:PLs}) as 
\begin{align}
\hat{I}_{\rm PL}^{\rm (n)}(s)
&\approx \left(1+\frac{k_{\rm q}R}{5D} \right)
 \left(s+k_{\rm t}+\frac{3k_{\rm q}}{R}
 \right)^{-1}.
\label{eq:PLsappr}
\end{align}
After the inverse Laplace transformation, we obtain
\begin{align}
I_{\rm PL}^{\rm (n)}(t)
&\approx \left(1+\frac{k_{\rm q}R}{5D} \right)\exp\left[
-\left(k_{\rm t} +\frac{3k_{\rm q}}{R}\right) t\right].
\label{eq:PLsappr1}
\end{align}
The rate constant of PL decay is inversely proportional to the particle radius apart from the natural decay rate constant given by $k_{\rm t}$. 
The PL-decay rate constant given by $k_{\rm t} +3k_{\rm q}/R$ is influenced by the quenching rate constant at the particle surface denoted by $k_{\rm q}$ and is not influenced by the diffusion coefficient in the limit of $R\sqrt{k_{\rm t}/D} <1$ at long times.
We also note that the PL quantum yield obtained by time integration of Eq. (\ref{eq:PLsappr1}) is less than one because $[R/(5D)]/[3/(Rk_{\rm t})]<1$ and $k_{\rm r} < k_{\rm t}$. 

The short-time kinetics can be obtained by assuming $R/L_{\rm D}(s)>1$. 
Notably, $R/L_{\rm D}(s)>1$ can be satisfied even when $R/L_{\rm D}(s=0)<1$. 
The results of short-time kinetics are valid for a certain time approximated by $t<1/(D/R^2-k_{\rm t})$.
Using $x \coth(x)-1 \approx x $ for $x>1$, we can approximate Eq. (\ref{eq:PLs}) as 
\begin{align}
\hat{I}_{\rm PL}^{\rm (n)}(s)
 &\approx  \frac{1}{s+k_{\rm t}}
\left[  1
 - 3 \frac{L_{\rm D}^2(s)}{R}
 \left(\frac{D}{k_{\rm q}}+
L_{\rm D}(s)
 \right)^{-1}
\right] 
\nonumber \\
&=\frac{1}{s+k_{\rm t}}
\left[  1
 -\frac{3 k_{\rm q}/R}{s+k_{\rm t}}
 \left(1+
\frac{k_{\rm q}}{D}\sqrt{\frac{D}{s+k_{\rm t}}}
 \right)^{-1}
\right].
\label{eq:PLsappr2}
\end{align}
The inverse Laplace transformation is naturally expressed using the 
two--parameter 
Mittag--Leffler function given by $E_{\alpha,\beta} (z)=\sum_{k=0}^\infty z^k/\Gamma(\beta+\alpha k)$, where ${\cal L}^{-1}\left(1/[s^2(1+h/\sqrt{s})]\right)=t E_{1/2,2} (-h \sqrt{t})$.\cite{Erdelyi_53}
Using Eq. (\ref{eq:PLsappr2}), we obtain  
\begin{align}
I_{\rm PL}^{\rm (n)}(t)
&\approx \exp\left(
-k_{\rm t}  t\right)\left(1-
\frac{3 k_{\rm q}}{R}\, t  E_{1/2,2} \left(-k_{\rm q} \sqrt{t/D} \right)\right).
\label{eq:PLsappr3}
\end{align}
Using the asymptotic expansion given by 
$E_{1/2,2} (z)\approx -2/\left(\sqrt{\pi} z \right)$ for $|\mbox{arg} z|=\pi$ in Eq. (\ref{eq:PLsappr3}), we find 
\begin{align}
I_{\rm PL}^{\rm (n)}(t)
&\approx \exp\left(
-k_{\rm t}  t\right)\left(1+(6/R)
\sqrt{Dt/\pi} \right), 
\label{eq:PLsappr4}
\end{align} 
where $k_{\rm q} \sqrt{t/D} >1$ is assumed. 
The condition of short-time kinetics can be expressed as $D/R^2-k_{\rm t}<1/t$, and the condition of asymptotic expansion is given by $k_{\rm q} \sqrt{t/D} >1$. 
Therefore, Eq. (\ref{eq:PLsappr4}) holds when $D(D/R^2-k_{\rm t} )< k_{\rm q}^2$. 
Otherwise, the short-time kinetics can be approximated by Eq. (\ref{eq:PLsappr3}). 

\subsubsection{Particle size larger than the diffusion length}
\label{sec:lparticlesize}

We here consider the opposite limit of $R/L_{\rm D}(s=0) >1$. 
If the particle radius is sufficiently large that $R\sqrt{k_{\rm t}/D} >1$ holds, then the results given by Eqs. (\ref{eq:PLsappr2})--(\ref{eq:PLsappr3}) approximately hold for all times because we have $D/R^2-k_{\rm t}<0<1/t$. 
The asymptotic kinetics is given by Eq. (\ref{eq:PLsappr4}). 

Equations (\ref{eq:PLsappr1}) and (\ref{eq:PLsappr4}) indicate that the long-time asymptotic decay obeys exponential kinetics apart from a weak time-dependent factor related to $\sqrt{Dt}/R$ in Eq. (\ref{eq:PLsappr4}). 
The rate constant of exponential decay in Eq. (\ref{eq:PLsappr1}) depends on the particle size $R$, whereas the rate constant of exponential decay in Eq. (\ref{eq:PLsappr4}) is independent of $R$. 

\subsection{Perfect quenching at the particle surface}
\label{sec:perfectq}

In the limit of $k_{\rm q} \rightarrow \infty$, Eq. (\ref{eq:PLs}) reduces to
\begin{align}
\hat{I}_{\rm PL}^{\rm (n)}(s)
&= \frac{1}{s+k_{\rm t}}
\left[  1
 - 3 \frac{L_{\rm D}(s)}{R}
  \left(\coth\left( \frac{R }{L_{\rm D}(s)}\right)- \frac{L_{\rm D}(s)} {R}\right)
\right].
\label{eq:PLs_pa}
\end{align}
We introduce the partial fraction expansion given by 
$\left(\coth(z)-1/z\right)/z=\sum_{j=1}^\infty 2/\left(z^2+(\pi j)^2\right)$
 and use 
${\cal L}^{-1}\left(1/[s(c s/r+1)]\right)=1-\exp\left(- r t/c\right)$
 to obtain 
the inverse Laplace transform of Eq. (\ref{eq:PLs_pa}) as\cite{NIST}
\begin{align}
I_{\rm PL}^{\rm (n)}(t)
=\sum_{n=1}^\infty \frac{6}{n^2 \pi^2} \exp \left(-k_{\rm t} t-\frac{Dn^2 \pi^2}{R^2} t
\right)
, 
\label{eq:PLs_pa_sol}
\end{align}
where $\sum_{n=1}^\infty 1/(n \pi)^2=1/6$ has been used.\cite{NIST} 
The asymptotic long-time decay can be approximated as 
\begin{align}
I_{\rm PL}^{\rm (n)}(t)
 \approx 
\frac{6}{\pi^2} \exp \left(-k_{\rm t} t-\frac{D\pi^2}{R^2} t
\right).
\label{eq:PLs_pa_solasym}
\end{align}
The long-time rate constant of PL decay contains the term inversely proportional to the square of the particle radius given by $D\pi^2/R^2$, which is in sharp contrast to 
the long-time PL-decay rate constant of $3k_{\rm q}/R$ given by Eq. (\ref{eq:PLsappr1}) when carriers are partially quenched at the particle surfaces. 
Table \ref{table:1} shows a summary of the limiting expressions of decay kinetics, together with the limiting expressions of quantum yield reported previously, where the underlying diffusion is assumed to be normal.\cite{Katoh_24}

\begin{table}[tb]
\extrarowheight=-3pt
\def\arraystretch{2}
\begin{tabular}{|c||c|c|}
\hline
& Partial surface quenching & Perfect surface quenching 
\\ \hline \hline
Quantum yield\cite{Katoh_24} &$\propto R$ & $\propto R^2$ \\ \hline
Long-time kinetics & 
$\displaystyle \exp\left[
-\left(k_{\rm t} +\frac{3k_{\rm q}}{R}\right) t\right]$
& $\displaystyle \exp \left[-\left( k_{\rm t} +\frac{D\pi^2}{R^2} \right) t
\right]$
\\ 
\hline
\end{tabular}
\caption{Influence of the surface quenching rate on the quantum yield and kinetics under normal diffusion 
when 
the particle size is smaller than the diffusion length ($R\sqrt{k_{\rm t}/D} <1$). 
$R$ indicates the particle radius, $k_{\rm t}$ is the inverse of the carrier bulk lifetime, the diffusion length is given by $\sqrt{D/k_{\rm t}}$, and $k_{\rm q}$ is the surface quenching rate constant. 
The limit of $k_{\rm q} \rightarrow \infty$ corresponds to the case of perfect surface quenching.  
The long-time kinetics is given essentially by an exponential decay for each case. 
The theoretical expressions of internal quantum yield have been reported elsewhere.\cite{Katoh_24} 
In the opposite limit of large particle size ($R\sqrt{k_{\rm t}/D} >1$), the long-time kinetics is given by $\displaystyle \exp\left(
-k_{\rm t}  t\right)$ and the quantum yield is given by $k_{\rm r}/k_{\rm t}$. 
}
\label{table:1}
\end{table}

\section{Theory: Subdiffusion}
\label{sec:subdiff}

Here, we consider the influence of subdiffusion kinetics on the PL decay. 
When carriers can be trapped in states with various energy levels, the carriers migrate via repeating trapping and detrapping processes. 
Once carriers are trapped in a trap state with a high detrapping energy barrier exceeding the thermal energy, detrapping hardly occurs. 
As time proceeds, the fraction of carriers trapped in trap states with energy barriers more than an order of magnitude greater than the thermal energy increases and the diffusion of carriers is slowed over time, particularly when the energy barrier is exponentially distributed.
The mean square displacement of carriers obeys $\langle r^2(t) \rangle =2 d D_\alpha t^\alpha$, and the exponent can be expressed as $\alpha =  k_{\rm B} T/E_0$ if the energy barrier height is exponentially distributed with the characteristic energy denoted by $E_0$.\cite{Scher_75,Pfister_78,Nelson_99,Barzykin_02,Seki_03_1,Seki_03,NELSON_04,Eaves_08,LIU_19}
The exponential density of states below the conduction band has been considered for charge carrier dynamics in TiO$_2$ nanoparticles.\cite{Nelson_99,Barzykin_02,NELSON_04,LIU_19}

If a self-trapped exciton dissociates into a mobile charge carrier and an immobile positive polaron, then repeated dissociation and association of a mobile charge carrier with immobile positive polarons can be regarded as subdiffusive migration of a self-trapped exciton when a self-trapped exciton is composed of a mobile charge carrier Coulombically bound to an immobile polaron with various configurations that yield an exponentially distributed binding energy. 
[Fig. \ref{Fig:model} (b)] 
This model is consistent with the observed red-shift of the PL emission wavelength over time after pulsed excitation. 

We solve the fractional diffusion equation describing subdiffusion extended to account for the natural decay rate constant ($k_{\rm t}$):\cite{Sokolov_06,Seki_07}
\begin{align}
\frac{\partial}{\partial t} p (r,t)=\frac{\partial}{\partial t}
D_{\alpha} \int_0^t d t_1 \frac{1}{\Gamma(\alpha)} \frac{\exp\left[- k_{\rm t} \left(t-t_1\right)\right]}{\left(t-t_1\right)^{1-\alpha}}\nabla^2 p (r,t_1)-k_{\rm t} p (r,t), 
\label{eq:diffds}
\end{align}
where the Laplacian is expressed by assuming spherical symmetry shown by Eq. (\ref{eq:laplace3d}).  
In the Laplace domain, Eq. (\ref{eq:diffds}) can be expressed as 
\begin{align}
s \hat{p} (r,s)-p(r,t=0)=D_{\alpha} \left(s+k_{\rm t} \right)^{1-\alpha} \nabla^2 \hat{p} (r,s)-k_{\rm t} \hat{p} (r,s). 
\label{eq:dispersivediffLap}
\end{align}

\subsection{Partial quenching at the particle surface}
\label{sec:partialq_sub}

We assume that carriers are partially quenched at the particle surface,\cite{Seki_03_1,Seki_03,Eaves_08}
\begin{align}
\left. - D_\alpha \frac{\partial}{\partial r} p (r,t)\right|_{r=R}= k_{{\rm q}\alpha} p (R,t).
\label{eq:BCds}
\end{align}
The boundary condition given by Eq. (\ref{eq:BCds}) has been derived for the case where the reaction at the boundary competes with the back transition of carriers from the surface to the bulk (inside the particle).\cite{Seki_03_1,Seki_03,Eaves_08}
Regarding the size dependence, Eq. (\ref{eq:BCds}) can be obtained as follows. 
By considering that the reflecting boundary condition should be expressed as 
$\left. - \left(s+k_{\rm t}\right)^{1-\alpha}D_\alpha \frac{\partial}{\partial r} \hat{p} (r,s)\right|_{r=R}= 0$, 
we can faithfully express the partial quenching boundary condition as  
$\left. - \left(s+k_{\rm t}\right)^{1-\alpha}D_\alpha \frac{\partial}{\partial r} \hat{p} (r,s)\right|_{r=R}= \left(s+k_{\rm t}\right)^{1-\alpha}k_{{\rm q}\alpha} \hat{p} (R,s)$. 
Therefore, the dimensionality of $k_{{\rm q}\alpha}$ differs from that of $k_{\rm q}$ of normal diffusion because the dimensionality of $D_{\alpha}$ differs from that of $D$.
Note that the dimensionality of $k_{{\rm q}\alpha}$ is [length]/[time]$^\alpha$ and the spatial dimensionality is still the same as that of $k_{\rm q}$. 
Also note that $k_{{\rm q}\alpha}$ is independent of $R$ for the same reason noted in the explanation of Eq. (\ref{eq:BC}).

We calculate the Laplace transform of the normalized PL intensity denoted by $\hat{I}_{\rm PL}^{\rm (n)} (s)$ as shown in Appendix C, 
where we have $I_{\rm PL}(0)=4\pi R^3 k_{\rm r} p_0/3$ by assuming uniform generation of excitons inside the spherical particle. 
We find for $I_{\rm PL}^{\rm (n)}(t)=I_{\rm PL}(t)/I_{\rm PL}(0)$, 
\begin{align}
\hat{I}_{\rm PL}^{\rm (n)}(s)
&= \frac{1}{s+k_{\rm t}}
\left[  1
 - 3 \frac{L_{{\rm D}\alpha}^2(s)}{R^2}
 \left(\frac{D_\alpha}{k_{{\rm q}\alpha}R}+
  \left(\frac{R}{L_{{\rm D}\alpha}(s)}\coth\left( \frac{R }{L_{{\rm D}\alpha}(s)}\right)- 1 \right)^{-1}
 \right)^{-1}
\right] ,
\label{eq:PLsds}
\end{align}
where $L_{{\rm D}\alpha}(s)$ is defined by 
\begin{align}
L_{{\rm D}\alpha}(s)=\sqrt{D_\alpha/(s+k_{\rm t})^\alpha}, 
\label{eq:LDsds}
\end{align}
which reduces to the diffusion length in the limit of $s\rightarrow 0$. 
\subsubsection{Particle size smaller than the diffusion length}
\label{sec:sub_particlesize_s}

First, we consider the case $R/L_{{\rm D}\alpha}(s=0)<1$, which can be expressed as $R\sqrt{k_{\rm t}^\alpha/D_\alpha} <1$; this condition applies when the particle size is sufficiently small. 

The asymptotic kinetics can be obtained by assuming $R/L_{{\rm D}\alpha}(s) <1$. 
Using $x \coth(x)-1 \approx x^2/3-x^4/(45) +\cdots $ for $x<1$, we can approximate 
Eq. (\ref{eq:PLsds}) 
as 
\begin{align}
\hat{I}_{\rm PL}^{\rm (n)}(s)
&\approx 
\frac{1}{s+k_{\rm t}}\left(1+\frac{3 k_{{\rm q}\alpha}L_{{\rm D}\alpha}(s)^2}{RD_\alpha}
\right)^{-1}
\nonumber \\
&\approx \left(1+ \frac{k_{{\rm q}\alpha}R}{5D_\alpha} \right)
 \left(s+k_{\rm t}+\frac{3k_{{\rm q}\alpha}}{R} \left(s+k_{\rm t}\right)^{1-\alpha}
 \right)^{-1}.
\label{eq:PLsapprds}
\end{align}
After the inverse Laplace transformation, we obtain 
\begin{align}
I_{\rm PL}^{\rm (n)}(t)
&\approx 
\left(1+ \frac{k_{{\rm q}\alpha}R}{5D_\alpha} \right)
\exp\left(-k_{\rm t} t\right)E_{\alpha} \left(-\frac{3k_{{\rm q}\alpha}}{R}t^\alpha\right),
\label{eq:PLsappr1ds}
\end{align}
where  $E_{\alpha} (z)=E_{\alpha,1} (z)=\sum_{k=0}^\infty z^k/\Gamma(\alpha k+1)$ is the one--parameter Mittag--Leffler function.\cite{Erdelyi_53}
The one--parameter Mittag--Leffler function can be regarded as the generalization of the exponential function ($E_1 (z)=\exp(z)$), 
where the one--parameter Mittag--Leffler function is an eigen function of a fractional-order time derivative in Caputo sense instead of the first-order derivative.\cite{Miller_93,Nagai_03,COFFEY_04}
In physical processes such as the dielectric relaxation, the one--parameter Mittag--Leffler function has been used to describe non-Markovian relaxation.\cite{COFFEY_04}  
The decay kinetics is characterized by $(3k_{{\rm q}\alpha}/R)t^\alpha$, which is not influenced by the diffusion coefficient in the limit of $R\sqrt{k_{\rm t}^\alpha/D_\alpha} <1$ at long times. 

Using\cite{Erdelyi_53},
\begin{align}
E_{\alpha} (-z^\alpha)&\approx 1-\frac{z^\alpha}{\Gamma(1+\alpha)}\approx \exp \left[- z^\alpha/\Gamma(1+\alpha)\right] \mbox{ for } 0<z<1
\label{eq:MLexpand} \\
E_{\alpha} (-z^\alpha)&\approx \frac{1}{\Gamma(1-\alpha)} \frac{1}{z^\alpha} \mbox{ for }  1>z, 
\label{eq:MLasymptotic}
\end{align}
we can approximate Eq. (\ref{eq:PLsappr1ds}) as
\begin{align}
I_{\rm PL}^{\rm (n)}(t)
&\approx \left(1+ \frac{k_{{\rm q}\alpha}R}{5D_\alpha} \right)
\exp\left[-k_{\rm t} t- \frac{3k_{{\rm q}\alpha}t ^\alpha}{\Gamma(1+\alpha)R}\right] \mbox{ for } t < \left(R/3k_{{\rm q}\alpha}\right)^{1/\alpha}
\label{eq:stretchedexp_v1}
\\
&\approx  \frac{\exp\left(-k_{\rm t} t \right)}{\Gamma(1-\alpha)}
\left(1+ \frac{k_{{\rm q}\alpha}R}{5D_\alpha} \right) \frac{R}{3k_{{\rm q}\alpha}t^\alpha} \mbox{ for } t > \left(R/3k_{{\rm q}\alpha}\right)^{1/\alpha}.
\label{eq:powerlaw_v1}
\end{align}  
When $k_{\rm t}$ can be ignored, Eq.  (\ref{eq:stretchedexp_v1}) represents the stretched (extended) exponential decay law with a size-dependent decay rate coefficient and Eq. (\ref{eq:powerlaw_v1}) represents the asymptotic power-law decay with a size-dependent amplitude.

The short-time kinetics can be obtained by assuming $R/L_{{\rm D}\alpha}(s)>1$. 
The short-time kinetics results are valid until a certain time approximated by $t<1/(D_\alpha^{1/\alpha}/R^{2/\alpha}-k_{\rm t})$.
Using $x \coth(x)-1 \approx x $ for $x>1$, Eq. (\ref{eq:PLs}) can be approximated as 
\begin{align}
\hat{I}_{\rm PL}^{\rm (n)}(s)
&\approx  \frac{1}{s+k_{\rm t}}
\left[  1
 - 3 \frac{L_{{\rm D}\alpha}^2(s)}{R}
 \left(\frac{D_\alpha}{k_{{\rm q}\alpha}}+
L_{{\rm D}\alpha}(s)
 \right)^{-1}
\right] 
\nonumber \\
&=\frac{1}{s+k_{\rm t}}
\left[  1
 -\frac{3 k_{{\rm q}\alpha}/R}{\left(s+k_{\rm t}\right)^\alpha}
\left(1+
\frac{k_{{\rm q}\alpha}}{D_\alpha}
\sqrt{\frac{D_\alpha}{\left(s+k_{\rm t}\right)^\alpha}}
 \right)^{-1}
\right].
\label{eq:PLsappr2ds}
\end{align}
The inverse-Laplace transformation is naturally expressed using the 
two--parameter 
Mittag--Leffler function given by 
$E_{\alpha,\beta} (z)=\sum_{k=0}^\infty z^k/\Gamma(\beta+\alpha k)$ as
${\cal L}^{-1}\left(1/[s^{1+\alpha}(1+h/\sqrt{s^\alpha})]\right)=t^\alpha E_{\alpha/2,1+\alpha} (-h \sqrt{t^\alpha})$. 
Note also that $k_{\rm t}$ in $s+k_{\rm t}$ indicates the Laplace transform of the function multiplied by $\exp(- k_{\rm t} t)$.
The asymptotic expansion is known; it is given by 
$E_{a,b} (z)\approx -1/\left(\Gamma(b-a) z \right)$ for $|\mbox{arg} z|=\pi$ and $0\leq a\leq1$, 
where $\Gamma(z)$ represents the Gamma function.\cite{Erdelyi_53}
Using Eq. (\ref{eq:PLsappr2ds}), we obtain  
\begin{align}
I_{\rm PL}^{\rm (n)}(t)
&\approx \exp\left(
-k_{\rm t}  t\right)\left(1-
\frac{3 k_{{\rm q}\alpha}}{R}\, t^\alpha  E_{\alpha/2,1+\alpha} \left(-k_{{\rm q}\alpha} \sqrt{t^\alpha/D_\alpha} \right)\right).
\label{eq:PLsappr3ds}
\end{align}
The asymptotic expansion of Eq. (\ref{eq:PLsappr3ds}) is obtained as 
\begin{align}
I_{\rm PL}^{\rm (n)}(t)
&\approx \exp\left(
-k_{\rm t}  t\right)\left(1+\frac{3}{R}
\frac{\sqrt{D_\alpha t^\alpha}}{\Gamma(1+\alpha/2)} \right), 
\label{eq:PLsappr4ds}
\end{align}
where $k_{{\rm q}\alpha} \sqrt{t^\alpha/D_\alpha} >1$ is assumed. 
Equation (\ref{eq:PLsappr4ds}) holds when $D_\alpha^{1/\alpha}(D_\alpha^{1/\alpha}/R^{2/\alpha}-k_{\rm t})< k_{{\rm q}\alpha}^{2/\alpha}$ 
by considering  $k_{{\rm q}\alpha} \sqrt{t^\alpha/D_\alpha} >1$ in addition to $D_\alpha^{1/\alpha}/R^{2/\alpha}-k_{\rm t}<1/t$. 
Otherwise, the short-time kinetics can be approximated by Eq. (\ref{eq:PLsappr3ds}). 

\subsubsection{Particle size larger than the diffusion length}
\label{sec:sub_particlesize_l}

We here consider the opposite limit of $R/L_{{\rm D}\alpha}(s=0)>1$. 
If the particle radius is sufficiently large that $R\sqrt{k_{\rm t}^\alpha/D_\alpha} >1$ holds, then the results given by Eqs. (\ref{eq:PLsappr2ds})--(\ref{eq:PLsappr3ds}) approximately hold for all times. 
The asymptotic kinetics can be approximated by Eq. (\ref{eq:PLsappr4ds}). 
Equation (\ref{eq:PLsappr4ds}) indicates an exponential decay with a weak correction factor related to $\sqrt{t^\alpha}$.

\subsection{Perfect quenching at the particle surface}
\label{sec:perfectq_alpha}

In the limit of $k_{\rm q} \rightarrow \infty$, Eq. (\ref{eq:PLsds}) reduces to
\begin{align}
\hat{I}_{\rm PL}^{\rm (n)}(s)
&= \frac{1}{s+k_{\rm t}}
\left[  1
 - 3 \frac{L_{{\rm D}\alpha}(s)}{R}
  \left(\coth\left( \frac{R }{L_{{\rm D}\alpha}(s)}\right)- \frac{L_{{\rm D}\alpha}(s)}{R} \right)
\right].
\label{eq:PLs_pa_alpha}
\end{align}
We introduce the partial fraction expansion given by 
$\left(\coth(z)-1/z\right)/z=\sum_{j=1}^\infty 2/\left(z^2+(\pi j)^2\right)$
and use
${\cal L}^{-1}\left(1/[s(c s^\alpha/r+1)]\right)=1-E_{\alpha}\left(
- r t^\alpha/c
\right)$ 
to obtain the inverse Laplace transform of Eq. (\ref{eq:PLs_pa_alpha}) as 
\begin{align}
I_{\rm PL}^{\rm (n)}(t)  &=\exp \left(-k_{\rm t} t\right)
\sum_{n=1}^\infty \frac{6}{n^2 \pi^2} 
E_{\alpha} \left(-
\frac{D_\alpha n^2 \pi^2}{R^2} t^\alpha
\right),
\label{eq:PLs_pa_sol_alpha}
\end{align}
where $\sum_{n=1}^\infty1/(n \pi)^2=1/6$ has been used.\cite{NIST}
The asymptotic long-time decay can be approximated as 
\begin{align}
I_{\rm PL}^{\rm (n)}(t)  \approx 
\frac{6}{\pi^2} \exp \left(-k_{\rm t} t\right)E_{\alpha} \left(-
\frac{D_\alpha \pi^2}{R^2} t^\alpha
\right).
\label{eq:PLs_pa_solasym_alpha}
\end{align}
The rate of PL decay is characterized by 
$\left[D_\alpha \pi^2/R^2\right]^{1/\alpha}$, 
which is a function of the square of the particle radius, in sharp contrast to the rate of PL decay characterized by $\left(3k_{{\rm q}\alpha}/R\right)^{1/\alpha}$ given by Eq. (\ref{eq:PLsappr1ds}) in the case where carriers are partially quenched at the particle surfaces.
Equation (\ref{eq:PLs_pa_solasym_alpha}) can be approximated as 
\begin{align}
I_{\rm PL}^{\rm (n)}(t)
 &\approx \frac{6}{\pi^2} 
\exp\left[-k_{\rm t} t- \frac{D_\alpha \pi^2}{R^2}t ^\alpha\right] \mbox{ for } t < \left[R^2/(D_\alpha \pi^2)\right]^{1/\alpha}
\label{eq:stretchedexp_pv1}
\\
&\approx  \frac{\exp\left(-k_{\rm t} t \right)}{\Gamma(1-\alpha)}
\frac{R^2}{D_\alpha \pi^2t^\alpha} \mbox{ for } t > \left[R^2/(D_\alpha \pi^2)\right]^{1/\alpha}.
\label{eq:powerlaw_pv1}
\end{align}
When $k_{\rm t}$ can be ignored, Eq.  (\ref{eq:stretchedexp_pv1}) represents the stretched (extended) exponential decay law with a size-dependent decay rate constant given by 
$\left[D_\alpha \pi^2/R^2\right]^{1/\alpha}$
and Eq. (\ref{eq:powerlaw_pv1}) represents the asymptotic power-law decay with a size-dependent amplitude.

\section{Photoluminescence quantum yield}
\label{sec:QY}

The PL quantum yield has already been studied under the assumption of normal diffusion.\cite{Katoh_24}
Here, we show that the size dependence of the PL quantum yield is essentially not altered by assuming subdiffusion inside the particle. 
We consider a spherical particle with radius $R$. 
We consider the case where $R$ is smaller than the absorption depth of incident light and assume a uniform distribution of carriers within the spherical particle. 
We denote the absorbed photon current density per unit volume by $g$. 
The absorption rate of irradiated light by a spherical particle with radius $R$ is given by $(4\pi/3) R^3 g$. 
We denote the steady-state distribution of carriers by 
$p_{\rm st} (r)$, where $r$ indicates the distance from the center of the particle. 
The PL rate in the steady state can be obtained from 
$I_{\rm PL}^{\rm (st)}=4\pi \int_0^R dr r^2 k_{\rm r} p_{\rm st} (r)$, where $ k_{\rm r}$ is the radiative decay rate constant. 
The internal quantum yield can be estimated from 
$\Phi_{\rm PL}=I_{\rm PL}^{\rm (st)}/[(4\pi/3) R^3 g]$. 
Equation (\ref{eq:dispersivediffLap}) can be regarded as the steady-state equation for 
$p_{\rm st}(r)$ by taking the limit of $s=0$ and regarding $\hat{p} (r,s=0)$ as 
$p_{\rm st}(r)$ and $p(r,t=0)$ as $g$. 
The mathematical equivalence is accompanied by the change in the dimensionality between $p(r,t=0)$ and $g$, 
where $g$ is the carrier density generated per unit time. 
According to the change in the dimensionality between $p(r,t=0)$ and $g$, 
$\hat{p} (r,s=0)$ can be regarded as $p_{\rm st}(r)$, though the dimensionality is changed.
By taking the limit and  
regarding $\hat{p} (r,s=0)$ as $p_{\rm st} (r)$ and $p(r,t=0)$ as $g$, we obtain
\begin{align}
\Phi_{\rm PL}=4\pi \int_0^R dr r^2 k_{\rm r} \hat{p} (r,s=0)/[(4\pi/3) R^3 p(r,t=0)] .
\label{eq:PhiPL}
\end{align}
Equation (\ref{eq:PhiPL}) corresponds to the kinetic measurement of the quantum yield.
The term $(4\pi/3) R^3 p(r,t=0)$ indicates the initial number of excitons inside the particle, whereas $\int_0^\infty dt k_{\rm r} 4\pi \int_0^R dr r^2 p(r,t)$ is the number of excitons that decay by PL. 
The latter quantity is equal to $4\pi \int_0^R dr r^2 k_{\rm r} \hat{p} (r,s=0)$ because $\hat{p} (r,s=0)=\int_0^\infty dt p(r,t)$. 

We have $\Phi_{\rm PL}=k_{\rm r}\hat{I}_{\rm PL}^{\rm (n)}(s=0)
$, and we obtain from Eq. (\ref{eq:PLsds})   
\begin{align}
\Phi_{\rm PL}&= \frac{k_{\rm r}}{k_{\rm t}}
\left[  1
 - 3 \frac{D_\alpha}{R^2 k_{\rm t}^\alpha}
 \left(\frac{D_\alpha}{Rk_{{\rm q}\alpha}}+
  \left(\frac{R}{\sqrt{D_\alpha/k_{\rm t}^\alpha}}\coth\left( \frac{R }{\sqrt{D_\alpha/k_{\rm t}^\alpha}}\right)- 1 \right)^{-1}
 \right)^{-1}
\right].
\label{eq:PhiPLsds}
\end{align}
Equation (\ref{eq:PhiPLsds}) represents the exact theoretical result of the internal quantum yield when the particle is uniformly excited.

\subsection{Partial quenching}
\label{sec:QYPartial}
\subsubsection{Particle size smaller than the diffusion length}
\label{sec:QYLarge}
First, we consider the case of $R\sqrt{k_{\rm t}^\alpha/D_\alpha} <1$; this condition applies when the particle size is sufficiently small. 

The asymptotic kinetics can be obtained by applying 
$x \coth(x)-1 \approx x^2/3-x^4/(45) +\cdots $ for $x<1$ to Eq. (\ref{eq:PhiPLsds}): 
\begin{align}
\Phi_{\rm PL}&\approx \frac{k_{\rm r}}{k_{\rm t}}\frac{Rk_{\rm t}^{\alpha}}{3k_{{\rm q}\alpha}}  \left(1+ \frac{k_{{\rm q}\alpha}R}{5D_\alpha} \right). 
\label{eq:PhiPLsds_1}
\end{align}
In this limit, $\Phi_{\rm PL} \propto R$ is obtained for the case of normal diffusion.\cite{Katoh_24}
Here, we show that $\Phi_{\rm PL} \propto R$ is obtained under the condition of  $k_{{\rm q}\alpha}R/(5D_\alpha)<1$; this condition is given by $k_{\rm q} R/(5D)<1$ for normal diffusion, and $\Phi_{\rm PL} \propto R^2$ is obtained by taking the limit of $k_{{\rm q}\alpha} \rightarrow \infty$.
Equation (\ref{eq:PhiPLsds_1}) is valid for the arbitrary value of $k_{{\rm q}\alpha}$. 
We will show that $k_{{\rm q}\alpha}$ can be determined by analyzing the PL kinetic data.
By knowing that $\Phi_{\rm PL} \propto R$, we can estimate the lower bound of $D_\alpha$ using $k_{{\rm q}\alpha}R/5<D_\alpha$. 
Measuring the internal quantum yield may be impractical. 
However, once the absolute value of $\Phi_{\rm PL}$ (the internal quantum yield) has been determined and $k_{{\rm q}\alpha}$ has been determined from the analysis of the PL kinetics, we can estimate $k_{\rm r}/k_{\rm t}$, which can be regarded as the quantum yield of excitons in the bulk in the absence of influence from the surfaces.
If we set $\alpha=1$ and $k_{{\rm q}\alpha} = k_{\rm q}$, then Eq. (\ref{eq:PhiPLsds_1}) reduces to the PL quantum yield under normal diffusion.

\subsubsection{Particle size larger than the diffusion length}
\label{sec:QYsmall}

Using $\coth(x)\approx 1 $ for $x>1$, we can approximate Eq. (\ref{eq:PhiPLsds}) as 
\begin{align}
\Phi_{\rm PL}&\approx    \frac{k_{\rm r}}{k_{\rm t}}
\left( 1 - \frac{3D_\alpha/\left(R k_{\rm t}^\alpha\right)}{{\sqrt{D_\alpha/k_{\rm t}^\alpha}}+D_\alpha /k_{{\rm q}\alpha}}
\right).
\label{eq:PLsappr2ds1}
\end{align}

\subsection{Quantum yield: Perfect quenching}

Using Eq. (\ref{eq:PhiPLsds}) and taking the limit of $k_{\rm q} \rightarrow \infty$, we obtain 
\begin{align}
\Phi_{\rm PL}&= \frac{k_{\rm r}}{k_{\rm t}}
\left[ 
1 - 3 \frac{\sqrt{D_\alpha/k_{\rm t}^\alpha}}{R}  \left(\coth\left( \frac{R }{\sqrt{D_\alpha/k_{\rm t}^\alpha}}\right)- \frac{\sqrt{D_\alpha/k_{\rm t}^\alpha}}{R} \right)
\right].
\label{eq:QYPLs_pa_alpha}
\end{align}

\subsubsection{Particle size smaller than the diffusion length}
\label{sec:PQYLarge}
Using $x \coth(x)-1 \approx x^2/3-x^4/45 \cdots$ for $x<1$, we can approximate Eq. (\ref{eq:QYPLs_pa_alpha}) as 
\begin{align}
\Phi_{\rm PL} &\approx 
\frac{1}{15} 
\frac{k_{\rm r}R^2}{k_{\rm t}^{1-\alpha} D_\alpha}.
\label{eq:QE2dsP}
\end{align}
Therefore, if the particle radius is smaller than the diffusion length ($R<L_{\rm D}$) under a perfectly quenching boundary condition at the particle surface, then the quantum efficiency is proportional to the square of the particle radius. 
Notably, Eq. (\ref{eq:PhiPLsds_1}) interpolates between the limit of $\Phi_{\rm PL} \propto R$ when $k_{{\rm q}\alpha}R/(5D_\alpha)<1$ and the limit of $\Phi_{\rm PL} \propto R^2$ of perfect quenching.

\subsubsection{Particle size larger than the diffusion length}

Using $\coth(x) \approx 1 $ for $x>1$, we can approximate Eq. (\ref{eq:QYPLs_pa_alpha}) as  
\begin{align}
\Phi_{\rm PL}&\approx   \frac{k_{\rm r}}{k_{\rm t}}
\left(
1 - 3 \frac{\sqrt{D_\alpha/k_{\rm t}^\alpha}}{R}
\right).
\label{eq:PLsappr2ds1P}
\end{align}
Equation (\ref{eq:PLsappr2ds1}) includes the limit of Eq. (\ref{eq:PLsappr2ds1P}) and the size dependence is not influenced by the value of $k_{{\rm q}\alpha}$, including the limit of $k_{{\rm q}\alpha} \rightarrow \infty$ (perfect quenching). 
In Table \ref{table:2}, we summarized the limiting expressions of long-time decay kinetics and internal quantum yields, where subdiffusion is assumed. 
If the carrier lifetime ($\tau$) inside the particle is defined by the argument of the one--parameter Mittag--Leffler function using $E_{\alpha} \left(-\left(t/\tau\right)^\alpha\right)$, we obtain $\tau \propto R^{1/\alpha}$ and $\tau \propto R^{2/\alpha}$ for the partial quenching case and the perfect quenching case, respectively. 
Particle-size-dependent lifetimes have been reported previously for rutile TiO$_2$, and an interpretation based on the current model is shown in Appendix D. 

\begin{table}[tb]
\extrarowheight=-3pt
\def\arraystretch{2}
\begin{tabular}{|c||c|c|}
\hline
&Partial surface quenching  & Perfect surface quenching 
\\
\hline \hline
Quantum yield & $\propto R$ & $\propto R^2$ \\ \hline
Long-time kinetics & 
$\displaystyle \exp\left(-k_{\rm t} t\right)E_{\alpha} \left(-\frac{3k_{{\rm q}\alpha}}{R}t^\alpha\right)
$
&
$\displaystyle \exp \left(-k_{\rm t} t\right)E_{\alpha} \left(-
\frac{D_\alpha \pi^2}{R^2} t^\alpha
\right)$
\\ \hline 
\end{tabular}
\caption{Influence of the surface quenching rate on the PL quantum yield and kinetics under subdiffusion 
when the particle size is smaller than the diffusion length 
($R\sqrt{k_{\rm t}^\alpha/D_\alpha} <1$). 
$R$ indicates the particle radius, $k_{\rm t}$ is the inverse of the bulk carrier lifetime, the diffusion length is given by $\sqrt{D_\alpha/k_{\rm t}^\alpha}$, and $k_{{\rm q}\alpha}$ is the surface quenching rate coefficient. 
Note that the dimensionality of $k_{{\rm q}\alpha}$ is [length]/[time]$^\alpha$ and the dimensionality of $D_\alpha$ is [length]$^2$/[time]$^\alpha$.
The limit of $k_{{\rm q}\alpha} \rightarrow \infty$ corresponds to the case of perfect surface quenching.  
$E_{\alpha} (z)=\sum_{k=0}^\infty z^k/\Gamma(\alpha k+1)$ is the one--parameter Mittag--Leffler function.\cite{Erdelyi_53}  [$E_1 (z)=\exp(z)$] 
The short-time expansion of $E_{\alpha} (-c t^\alpha)$ is the stretched (extended) exponential decay given by $ \exp \left[- c t^\alpha/\Gamma(1+\alpha)\right]$, where $c$ is a constant. 
The long-time asymptotic of $E_{\alpha} (-c t^\alpha)$ is the power-law decay given by $1/(\Gamma(1-\alpha)c t^\alpha)$. 
If the carrier lifetime inside the particle is defined using $(t/\tau)^\alpha$ in the one--parameter Mittag--Leffler function, we obtain $\tau \propto R^{1/\alpha}$ and $\tau \propto R^{2/\alpha}$ for the partial quenching case and the perfect quenching case, respectively. 
In the opposite limit of large particle size ($R\sqrt{k_{\rm t}^\alpha/D_\alpha}>1$), the long-time kinetics is given by $\displaystyle \exp\left(
-k_{\rm t}  t\right)$ and the quantum yield is given by $k_{\rm r}/k_{\rm t}$. 
}
\label{table:2}
\end{table}

\section{Discussion}
\label{sec:Discussion} 

\begin{figure}[h]
\begin{center}
\includegraphics[width=1.0\textwidth]{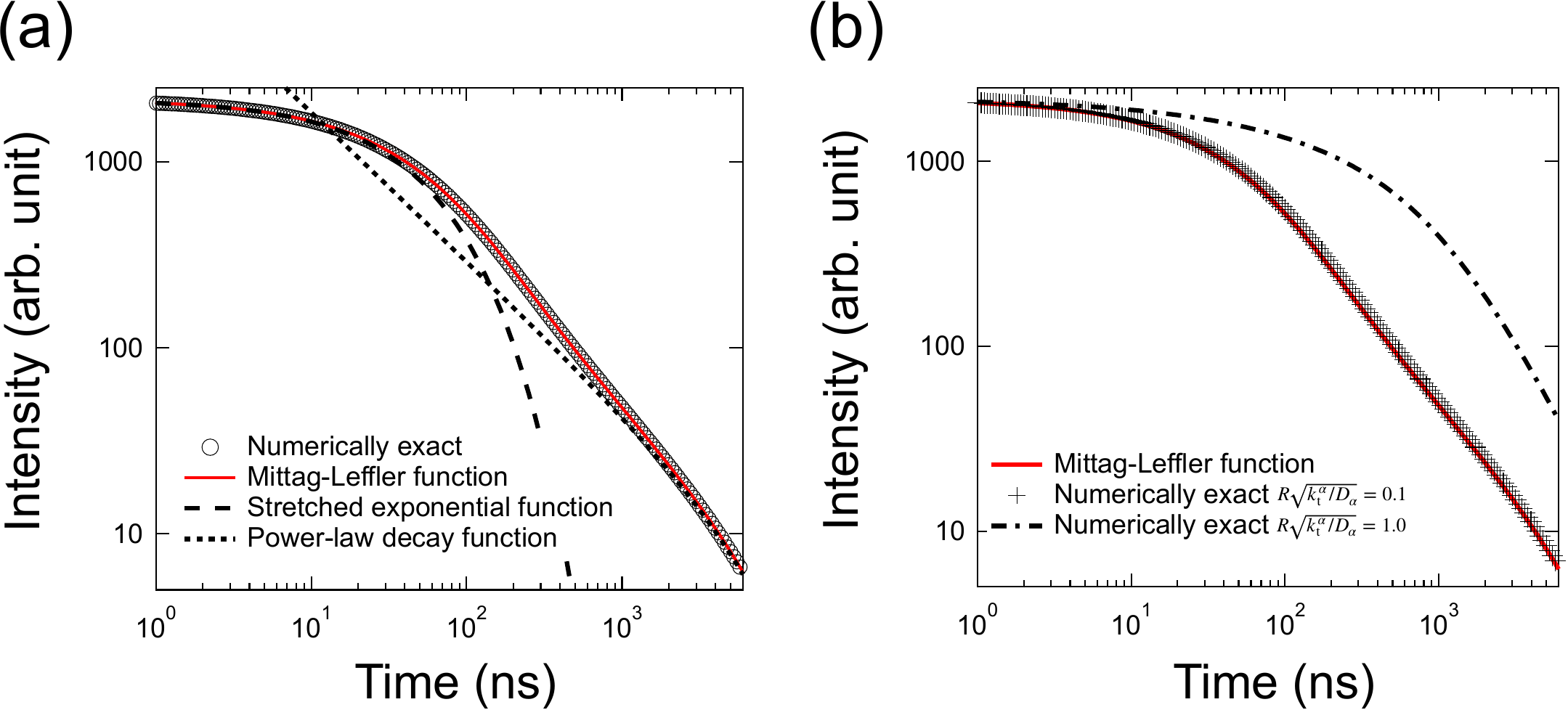}
\end{center}
\caption{(Color online) 
(a)
Circles represent the numerical inverse Laplace transform of Eq. (\ref{eq:PLsds}).\cite{Stehfest1970_47,Stehfest1970_624}
The (red) solid line indicates the decay expressed by the one--parameter Mittag--Leffler function [Eq. (\ref{eq:PLsappr1ds})].
The long-dashed line indicates the decay expressed by the stretched exponential decay function [Eq. (\ref{eq:stretchedexp_v1})].
The dots represent the decay expressed by the power-law decay function [Eq. (\ref{eq:powerlaw_v1})]. 
The parameters are the same as those corresponding to Fig. \ref{Fig:A200} [$k_{\rm t}=10^5$ s$^{-1}$, $\alpha=0.8$,  $k_{{\rm q}\alpha}=0.0214$ m s$^{-\alpha}$, and the value of $D_\alpha$ satisfying $R\sqrt{k_{\rm t}^\alpha/D_\alpha} =0.01$ with $R=100$ nm.]
(b) 
The + symbols and the dash-dotted line represent the numerical inverse Laplace transform of Eq. (\ref{eq:PLsds}) for $R\sqrt{k_{\rm t}^\alpha/D_\alpha} =0.1$ and $R\sqrt{k_{\rm t}^\alpha/D_\alpha} =1$ with $R=100$ nm, respectively.\cite{Stehfest1970_47,Stehfest1970_624}
The (red) solid line indicates the decay expressed by the one--parameter Mittag--Leffler function [Eq. (\ref{eq:PLsappr1ds})].
The parameters other than $D_\alpha$ are the same as those corresponding to subfigure (a). 
Equation (\ref{eq:PLsappr1ds}) is valid for $t\geq 1$ ns when $R\sqrt{k_{\rm t}^\alpha/D_\alpha} <1$. 
}
\label{Fig:theory1}
\end{figure}
We have analyzed the PL decay kinetics of anatase TiO$_2$ photocatalysts using the one--parameter Mittag--Leffler function [Eq. (\ref{eq:PLsappr1ds})], which is an approximate solution of the model. 
In this section, we justify Eq. (\ref{eq:PLsappr1ds}) by comparison with the numerical exact solution of the model for the parameters used to draw Fig. \ref{Fig:A200}. 
The exact solution of the model is shown in the Laplace space by Eq. (\ref{eq:PLsds}). 
The numerically exact decay kinetics is obtained via the numerical inverse Laplace transform of Eq. (\ref{eq:PLsds}).\cite{Stehfest1970_47,Stehfest1970_624}

We have theoretically shown that the decay expressed using the one--parameter Mittag--Leffler function approximates the exact solution for $t >1/(D_\alpha^{1/\alpha}/R^{2/\alpha}-k_{\rm t})$
when the condition given by $R\sqrt{k_{\rm t}^\alpha/D_\alpha} <1$ is satisfied. 
The condition $t >1/(D_\alpha^{1/\alpha}/R^{2/\alpha}-k_{\rm t})$ ($t>30$ ns) was 
roughly estimated  to derive Eq. (\ref{eq:PLsappr1ds}). 
As shown in Fig. \ref{Fig:theory1}, the decay expressed using the one--parameter Mittag--Leffler function reproduces the exact numerical decay function when $R\sqrt{k_{\rm t}^\alpha/D_\alpha} <1$ for $t\geq 1$ ns. 
Figure \ref{Fig:theory1} indicates that the estimation is too rough and the lower bound can be ignored for the time range of $t\geq 1$ ns.  

The short-time kinetics can be further approximated using the stretched exponential function [Eq. (\ref{eq:stretchedexp_v1})]. 
The long-time asymptotic decay can be approximated using the power-law decay function [Eq. (\ref{eq:powerlaw_v1})]. 
The characteristic time dividing the two time regimes denoted by $\tau$ is obtained from $\tau=\left[ R/(3k_{{\rm q}\alpha}) \right]^{1/\alpha}$; $\tau$ is the decay constant used in $E_{\alpha} \left[-(t/\tau)^\alpha \right]$ obtained from Eq. (\ref{eq:PLsappr1ds}). 
The characteristic time can be estimated as $\tau=55$ ns using $k_{{\rm q}\alpha}=0.0214$ m s$^{-\alpha}$, $R=100$ nm, and $\alpha=0.8$. 
According to Fig. \ref{Fig:theory1}(a), Eq. (\ref{eq:stretchedexp_v1}) approximates the exact decay kinetics when $t \leq \tau$, whereas Eq. (\ref{eq:powerlaw_v1}) approximates the exact decay kinetics when $t > 10\tau$. 

The expression of the decay kinetics given by the 
one--parameter 
Mittag--Leffler function [Eq. (\ref{eq:PLsappr1ds})] is independent of the value of the diffusion coefficient ($D_\alpha$) because 
the one--parameter Mittag--Leffler function is obtained when the decay is limited by the reaction rate of surface quenching. 
The one--parameter Mittag--Leffler function is highly nonexponential when $\alpha$ is less than $1$; nevertheless, it is derived under the reaction-limited condition. 

When the condition given by $R\sqrt{k_{\rm t}^\alpha/D_\alpha} < 1$ is satisfied, the one--parameter Mittag--Leffler function [Eq. (\ref{eq:PLsappr1ds})] approximates the exact solution obtained from Eq. (\ref{eq:PLsds}) and the decay kinetics should not depend on the value of the diffusion coefficient $D_\alpha$; $D_\alpha$ cannot be estimated using the kinetics. 
Only the lower bound of $D_\alpha$ might be estimated. 
Fig. \ref{Fig:theory1}(b) shows that the one--parameter Mittag--Leffler function [Eq. (\ref{eq:PLsappr1ds})] approximates the exact solution when $R\sqrt{k_{\rm t}^\alpha/D_\alpha} =0.1$ and fails to approximate the exact solution when $R\sqrt{k_{\rm t}^\alpha/D_\alpha} =1$. 
Using $R\sqrt{k_{\rm t}^\alpha/D_\alpha} \leq 0.1$, the lower bound of the diffusion coefficient of subdiffusion can be estimated as $D_\alpha=1 \times 10^{-8}$ m$^2$ s$^{-\alpha}$, where $k_{\rm t}=10^5$ s$^{-1}$, $\alpha=0.8$, and $R=100$ nm are used. 
The lower bound, however, depends on the accuracy of the value of $k_{\rm t}$. 
Parameter $k_{\rm t}$ is the inverse of the bulk carrier lifetime, and $k_{\rm t}=10^5$ s$^{-1}$ can be the largest value of $k_{\rm t}$ according to our data up to $4~\mu$s. 
If the value of $k_{\rm t}$ decreases, the lower bound of $D_\alpha$ can be further decreased.  

\section{Conclusion}
\label{sec:Conclusion} 

We have extended a theoretical model introduced to explain the size dependence of the PL quantum yield to study the nonexponential decay kinetics of PL.\cite{Katoh_24}
Previously, we theoretically showed that the PL quantum yield is proportional to the particle size when the diffusion length exceeds the particle size and the PL is limited by the rate constant of quenching on the surface of the particle by assuming normal diffusion. 
However, highly nonexponential decay of PL cannot be obtained if we assume normal diffusion in the bulk (Table \ref{table:1}). 
The model is extended to consider subdiffusion rather than normal diffusion and thereby be consistent with the observed complex decay kinetics. 

Table \ref{table:2} summarizes the limiting expressions of long-time decay kinetics and internal quantum yield, where subdiffusion is assumed. 
When the diffusion length, which can be properly defined for the case of subdiffusion, exceeds the particle size and carriers are partially quenched at the particle surface, the PL decay can be approximately expressed using 
the one--parameter Mittag--Leffler function [Eq. (\ref{eq:PLsappr1ds})]. 
The one--parameter Mittag--Leffler function has been used to describe non-Markovian relaxation.\cite{COFFEY_04}  
The expression of the decay kinetics given by the one--parameter Mittag--Leffler function [Eq. (\ref{eq:PLsappr1ds})] is independent of the value of the diffusion coefficient $D_\alpha$. 
The value of $D_\alpha$ cannot be estimated using the kinetics; only the lower bound can be estimated. 
The decay expressed using the one--parameter Mittag--Leffler function can be further approximated using a stretched exponential function [Eq. (\ref{eq:stretchedexp_v1})] at short times. 
The long-time asymptotic decay can be approximated using a power-law decay function [Eq. (\ref{eq:powerlaw_v1})]. 
The characteristic time dividing the two time regimes denoted by $\tau$ is obtained from $\tau=\left[ R/(3k_{{\rm q}\alpha}) \right]^{1/\alpha}$; although the decay is nonexponential, $\tau$ essentially represents the PL decay time constant, where PL decays by visiting quenching sites on the particle surface. 
If the subdiffusion property ($\alpha$) and surface quenching rate coefficient ($k_{{\rm q}\alpha}$) are not altered by a change of the particle size, the present model based on the partial quenching at the particle surfaces can be justified by studying the particle-size dependence of the characteristic time given by $\tau=\left[ R/(3k_{{\rm q}\alpha}) \right]^{1/\alpha}$ (Appendix D). 
When $\alpha$ changes dramatically as the particle size is varied, the bulk transport property inside the particle might be modified. 

If a self-trapped exciton dissociates into a mobile charge carrier and an immobile positive polaron, repeated dissociation and association of a mobile charge carrier with immobile positive polarons forming a self-trapped exciton of various configurations can be naturally regarded as subdiffusive migration of a self-trapped exciton.  
This model is consistent with the observed shift in the PL spectrum toward lower energy with time after the pulsed excitation. 
Our theoretical analysis of the PL quantum yield and the PL decay kinetics provides a comprehensive picture of mobile charge carriers, immobile polarons, and self-trapped excitons. 

In the model, diffusing carriers can be either excitons or mobile charge carriers.\cite{TANG_93,Katoh_22,Katoh_24,Nandal_21} 
In addition, similar models that consider only normal diffusion have been studied for different observables.\cite{Thiele_39,Aris_75,Carslaw_86,Mitra_92,PRICE_98}
The proposed theoretical model has a wide range of applications.

\acknowledgments
This work was supported by a Grant-in-Aid for Scientific Research (No. 20H02699) from the Ministry of Education, Culture, Sports, Science and Technology of Japan.

\section*{Data Availability Statement}
The data that support the findings of this study are available from the corresponding author upon reasonable request. 

\newpage
\setcounter{equation}{0}  
\section*{Appendix A. PL quantum yield against particle size: Anatase TiO$_2$}
\renewcommand{\theequation}{A.\arabic{equation}}  
\setcounter{equation}{0}  
\renewcommand{\thefigure}{A.\arabic{figure}}  
\setcounter{figure}{0}  

\begin{figure}[h]
\begin{center}
\includegraphics[width=0.5\textwidth]{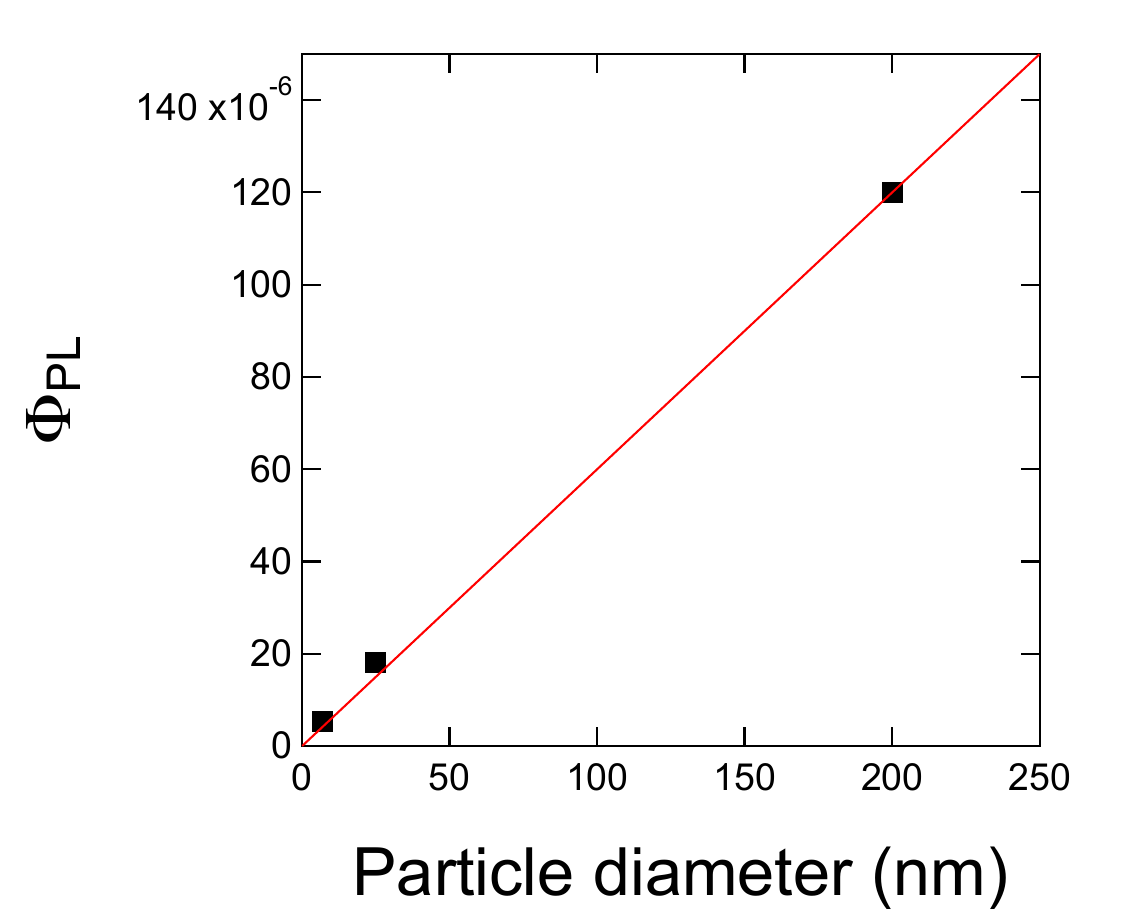}
\end{center}
\caption{(Color online) 
PL quantum yields for three anatase TiO$_2$ photocatalysts are plotted as functions of particle diameter (7, 25, and 200 nm for A-7, A-25, and A-200, respectively).\cite{Katoh_24}
The quantum yield was obtained under steady-state (continuous wave) excitation conditions. 
The (red) line indicates the proportional relation obtained by fitting to the data for particle diameters of 7 and 200 nm and the origin. 
The proportional constant is $6.0 \times 10^{-7}$ m$^{-1}$. 
}
\label{Fig:QE}
\end{figure}

\renewcommand{\theequation}{B.\arabic{equation}} 
\setcounter{equation}{0}  
\section*{Appendix B. Derivation of Eq. (\ref{eq:PLs})}
We introduce $\hat{Z}(r,s)=r \hat{p} (r,s)$. 
By substituting $ \hat{p} (r,s)=\hat{Z}(r,s)/r$ into the Laplace transform of Eq. (\ref{eq:diff}) with Eq. (\ref{eq:laplace3d}), we obtain,
\begin{align}
0= \frac{\partial^2}{\partial r^2} \hat{Z}(r,s)-\frac{s+k_{\rm t}}{D} \hat{Z}(r,s)+\frac{p(r,0) r}{D}.
\label{eq:diff_g_2}
\end{align}
The solution can be expressed as 
\begin{align}
\hat{Z}(r,s)=A \sinh \left(\frac{r}{L_{\rm D}(s)}\right)+B \cosh \left(\frac{r}{L_{\rm D}(s)} \right) +C_0 r + C_1,
\label{eq:Z_1}
\end{align}
where $L_{\rm D}(s)$ is defined by Eq. (\ref{eq:LDs}).

By substituting Eq. (\ref{eq:Z_1}) into Eq. (\ref{eq:diff_g_2}), we find $\left(s+k_{\rm t}\right) (C_0 r+C_1)=p(r,0) r$ and obtain $C_1=0$ and $C_0=p(r,0)/\left(s+k_{\rm t}\right)$.
Because $p (r,t)$ should be finite, we have $B=0$. 
The solution of Eq. (\ref{eq:diff}) can be rewritten as 
\begin{align}
\hat{p} (r,s)=\frac{A}{r} \sinh \left(\frac{r}{L_{\rm D}(s)} \right) +\frac{p(r,0) }{s+k_{\rm t}}. 
\label{eq:pm_1}
\end{align}

Using the boundary condition given by Eq. (\ref{eq:BC}), we obtain
\begin{align}
-\frac{Ak_{\rm q}}{R} \sinh \left(\frac{R}{L_{\rm D}(s)} \right) \left[1+\frac{D}{Rk_{\rm q}}\left(\frac{R}{L_{\rm D}(s)} \coth \left(\frac{R}{L_{\rm D}(s)} \right)-1\right)\right]=\frac{k_{\rm q}p(r,0) }{s+k_{\rm t}}.
\label{eq:pm_A}
\end{align}
Finally, we obtain 
\begin{align}
\hat{p} (r,s)=\frac{p(r,0) }{s+k_{\rm t}}
\left[1-
\frac{R}{r} \frac{ \sinh \left(r /L_{\rm D}(s)\right)}{ \sinh \left(R /L_{\rm D}(s) \right) }
\left(1+\frac{D}{Rk_{\rm q}}
\left(\frac{R}{L_{\rm D}(s)} \coth \left(\frac{R }{L_{\rm D}(s)} \right)-1
\right)\right)^{-1}
\right].
\label{eq:pm_2}
\end{align}
Using $p(r,0)=p_0$, we find Eq. (\ref{eq:PLs}).

\renewcommand{\theequation}{C.\arabic{equation}}  
\setcounter{equation}{0}  
\section*{Appendix C. Derivation of Eq. (\ref{eq:PLsds})}

We introduce $\hat{Z}(r,s)=r \hat{p} (r,s)$ and obtain, from Eq. (\ref{eq:diffds}),  
\begin{align}
0= \frac{\partial^2}{\partial r^2} \hat{Z}(r,s)-\frac{\left(s+k_{\rm t}\right)^\alpha}{D_\alpha} \hat{Z}(r,s)+\frac{p(r,0) r}{D_\alpha \left(s+k_{\rm t} \right)^{1-\alpha}} .
\label{eq:diff_g_2ds}
\end{align}
The solution can be expressed as 
\begin{align}
\hat{Z}(r,s)=A_\alpha \sinh \left(\frac{r}{L_{{\rm D}\alpha}(s)} \right)+B_\alpha \cosh \left(\frac{r}{L_{{\rm D}\alpha}(s)}\right) +C_0 r + C_1 ,
\label{eq:Z_1ds}
\end{align}
where $L_{{\rm D}\alpha}(s)$ is defined by Eq. (\ref{eq:LDsds}). 

By substituting Eq. (\ref{eq:Z_1ds}) into Eq. (\ref{eq:diff_g_2ds}), we find $\left(s+k_{\rm t}\right) (C_0 r+C_1)=p(r,0) r$ and obtain $C_1=0$ and $C_0=p(r,0) /\left(s+k_{\rm t}\right)$.
Because $p (r,t)$ should be finite, we have $B_\alpha=0$. 
The solution of Eq. (\ref{eq:diff}) can be rewritten as 
\begin{align}
\hat{p} (r,s)=\frac{A_\alpha}{r} \sinh \left(\frac{r}{L_{{\rm D}\alpha}(s)}\right) +\frac{p(r,0) }{s+k_{\rm t}}    .
\label{eq:pm_1ds}
\end{align}

Using the boundary condition given by Eq. (\ref{eq:BC}), we obtain
\begin{align}
-\frac{A_\alpha k_{{\rm q}\alpha}}{R} \sinh \left(\frac{R}{L_{{\rm D}\alpha}(s)}  \right) 
\left[1+\frac{D_\alpha}{Rk_{{\rm q}\alpha}}
\left(\frac{R}{L_{{\rm D}\alpha}(s)}\, 
\coth \left(\frac{R}{L_{{\rm D}\alpha}(s)} \right)-1\right)
\right]
=\frac{k_{{\rm q}\alpha}p(r,0) }{s+k_{\rm t}}.
\label{eq:pm_Ads}
\end{align}

Finally, we obtain 
\begin{multline}
\hat{p} (r,s)=
\\
\frac{p(r,0) }{s+k_{\rm t}}
\left[1-
\frac{R}{r} \frac{ \sinh \left(r /L_{{\rm D}\alpha}(s)\right)}{ \sinh \left(R /L_{{\rm D}\alpha}(s) \right) }
\left(1+\frac{D_\alpha}{Rk_{{\rm q}\alpha}}
\left(\frac{R}{L_{{\rm D}\alpha}(s)} \coth \left(\frac{R }{L_{{\rm D}\alpha}(s)} \right)-1
\right)\right)^{-1}
\right]. 
\label{eq:pm_2ds}
\end{multline}
Using $p(r,0)=p_0$, we find Eq. (\ref{eq:PLsds}).

\renewcommand{\theequation}{D.\arabic{equation}}  
\setcounter{equation}{0}  
\section*{Appendix D. PL lifetime as a function of particle size: Rutile TiO$_2$}
\renewcommand{\theequation}{D.\arabic{equation}}  
\setcounter{equation}{0}  
\renewcommand{\thefigure}{D.\arabic{figure}}  
\setcounter{figure}{0}  
\begin{figure}[h]
\begin{center}
\includegraphics[width=0.5\textwidth]{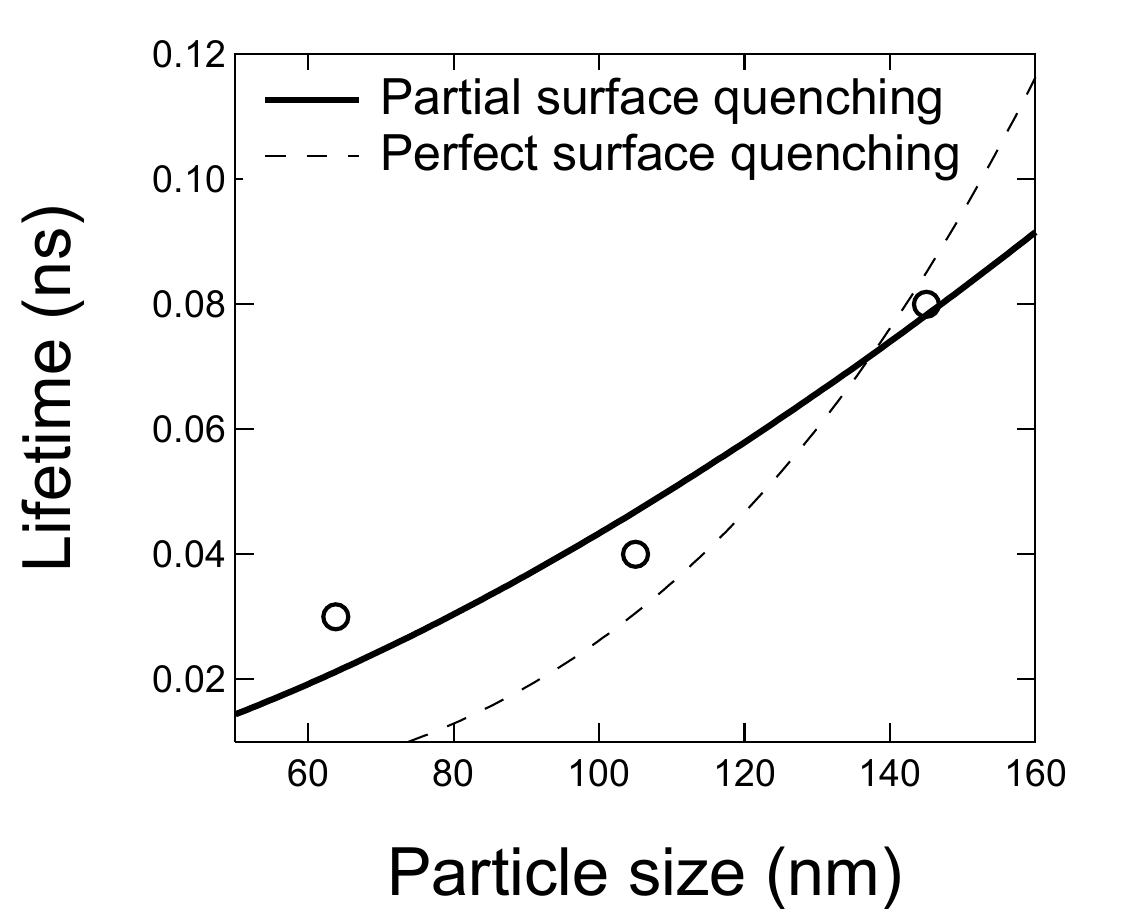}
\end{center}
\caption{
The PL lifetimes of three rutile TiO$_2$ photocatalysts are shown as functions of their particle size.
The black circles indicate the published experimental data (PT101 of Table 1 of Ref. \onlinecite{FUJIHARA_00}).  
The thick solid line and thin dashed line indicate $\tau \propto R^{1/\alpha}$ (partial surface quenching case) and $\tau \propto R^{2/\alpha}$ (perfect surface quenching case) with $\alpha=0.63$, respectively; $\alpha=0.63$ is shown in Table 1 of Ref. \onlinecite{FUJIHARA_00}, where a modified stretched exponential function is used to estimate $\alpha$. 
 }
\label{Fig:Rutile}
\end{figure}

The PL decay kinetics of rutile TiO$_2$ photocatalysts obtained from Ishihara (PT101) were measured by applying pulsed photoexcitation at 355 nm (0.22 mJ pulse$^{-1}$); PL was obtained at 480 nm.\cite{FUJIHARA_00}
The decay profile was analyzed using a modified stretched exponential function ($t^{\alpha-1} \exp \left[-(t/\tau)^{\alpha} \right]$), where $\tau$ is the PL lifetime of the particulate photocatalysts and $\alpha$ is a fitting parameter.\cite{FUJIHARA_00}
As shown in Table \ref{table:2}, we derived $\tau \propto R^{1/\alpha}$ and $\tau \propto R^{2/\alpha}$ for the partial quenching case and the perfect quenching case, respectively, when $\alpha$ and $k_{{\rm q}\alpha}$ are size-independent. 
As shown in Table 1 of Ref. \onlinecite{FUJIHARA_00}, $\alpha$ was indeed size-independent. 

The published data for rutile TiO$_2$ (Table 1 of Ref. \onlinecite{FUJIHARA_00}) are analyzed using the relation of $\tau \propto R^{1/\alpha}$ or $\tau \propto R^{2/\alpha}$ in Fig. \ref{Fig:Rutile}. 
Previously, the migration of carriers from the bulk of the particles to the surface was concluded from the observation of increasing lifetime of the emission with increasing particle size.\cite{FUJIHARA_00}
Our results give a more concrete picture that the PL lifetime of particulate photocatalysts depends on the particle size as a result of partial quenching at the particle surface; the exponent $\alpha$ might reflect the subdiffusive motion in the bulk, which should not be influenced by the particle size in the current model. 
However, the above analysis might be too naive because the stretched exponential function is a limited form of the more accurate expression given by the one--parameter Mittag--Leffler function [Eq. (\ref{eq:PLsappr1ds})]. 
Here, we include the analysis for illustrating a theoretical method to analyze the PL decay rate as a function of particle size.

\newpage
%

\end{document}